\documentclass{article}

\usepackage{arxiv}

\usepackage[utf8]{inputenc} 
\usepackage[T1]{fontenc}    
\usepackage{hyperref}       
\usepackage{url}            
\usepackage{booktabs}       
\usepackage{amsfonts}       
\usepackage{nicefrac}       
\usepackage{microtype}      
\usepackage{lipsum}
\usepackage{graphicx}
\usepackage{natbib}
\usepackage{booktabs}
\usepackage{caption}
\usepackage{multirow}
\usepackage{makecell}
\usepackage{textgreek}
\usepackage{amsmath}
\usepackage{amsthm}
\usepackage{graphicx}
\usepackage{rotating}
\usepackage{makecell}
\usepackage{array}
\usepackage{float}
\usepackage{lscape}
\usepackage{hyperref}
\usepackage{cleveref}
\usepackage{caption}
\usepackage{algorithm}
\usepackage{algpseudocode}
\usepackage{natbib}
\usepackage{url}
\usepackage{longtable}

\title{XDementNET: An Explainable Attention Based Deep Convolutional Network to Detect Alzheimer Progression from MRI data}

\author{
 Soyabul Islam Lincoln \\
 Dept. of Electronics and Communication Engineering\\
 Khulna University of Engineering \& Technology\\
 KUET Road, Fulbarigate, Khulna-9203.\\
\texttt{islam1709028@stud.kuet.ac.bd} \\
   \And
 Mirza Mohd Shahriar Maswood\\
 Dept. of Electronics and Communication Engineering\\
 Khulna University of Engineering \& Technology\\
 KUET Road, Fulbarigate, Khulna-9203.\\
  \texttt{shahriar@ece.kuet.ac.bd} \\
  }

\begin{document}
\maketitle
\begin{abstract}
A common neurodegenerative disease, Alzheimer's disease requires a precise diagnosis and efficient treatment, 
particularly in light of escalating healthcare expenses and the expanding use of artificial intelligence in medical diagnostics.
Many recent studies shows that the combination 
of brain Magnetic Resonance Imaging (MRI) and deep neural networks have achieved promising results for diagnosing AD. 
Using deep convolutional neural networks, this paper introduces a novel deep learning architecture that 
incorporates multiresidual blocks, specialized spatial attention blocks, grouped query attention, and
multi-head attention. The study assessed the model's performance on four publicly accessible datasets 
and concentrated on identifying binary and multiclass issues across various categories.
This paper also takes into account of the explainability of AD's progression and compared with 
state-of-the-art methods namely Gradient Class Activation Mapping (GradCAM),  Score-CAM, Faster 
Score-CAM, and XGRADCAM. Our methodology consistently outperforms current approaches, achieving 99.66\% accuracy
in 4-class classification, 99.63\% in 3-class classification, and 100\% in binary classification using Kaggle datasets.
For Open Access Series of Imaging Studies (OASIS) datasets the accuracies are 99.92\%, 99.90\%, and 99.95\% respectively. 
The Alzheimer's Disease Neuroimaging Initiative-1 (ADNI-1) dataset was used for experiments in three planes (axial, sagittal, and coronal) and a combination of all planes.
The study achieved accuracies of 99.08\% for axis, 99.85\% for sagittal, 99.5\% for coronal, and 99.17\% 
for all axis, and 97.79\% and 8.60\% respectively for ADNI-2. The network's ability to retrieve important information from MRI images is demonstrated by its excellent accuracy in categorizing AD stages.

\end{abstract}


\section{Introduction}\label{intro}
Throughout life, the human brain experiences intricate, age-related changes. Throughout 
development, these alterations exhibit non-linear and region-specific patterns. Cognitive 
function often deteriorates with age, resulting in widespread brain atrophy \cite{cole2017predicting}. 
As humans age, brain tissue and nerve cells die gradually by pairs of TAU proteins that 
stabilize the microtubules, leading to Alzheimer's disease. Some symptoms are 
memory loss, wording, confounding, and disorientation. In 2018, there were almost 50 million 
victims worldwide, extending the number to 131 million by 
2050, and the semi-economic cost will be 9.12 trillion dollars \citep{scheltens2021alzheimer}.  
Although AD typically affects people 65 and older, the number of cases with early 
onset suggests that it cannot be purely categorized as an old age disease \citep{alzheimer20192019}. 
Unfortunately, AD has no known cure, and the medicines that are now available merely slow down the disease's course \citep{neugroschl2011alzheimer}. 
Although AD can develop in some persons with mild cognitive impairment (MCI), not all 
MCI patients go on to acquire AD. The diagnosis of brain disorders relies heavily on a 
number of neuroimaging methods, including Computed Tomography (CT), Positron Emission 
Tomography (PET), Magnetic Resonance Imaging (MRI), and Functional Magnetic Resonance 
Imaging (fMRI).The great majority of AD diagnosis methods now in use depend on 
laborious manual assessment by medical experts. 
In order to reduce medical costs, enhance treatment results, and postpone the abnormal 
degeneration of the brain, researchers are exploring early diagnosis for this disorder \citep{wu2022attention}. 
Modern brain imaging techniques like Positron Emission Tomography (PET) 
and Magnetic Resonance Imaging (MRI) can be used to identify AD early on. 
Clinically, MRI is widely used for AD-related diagnosis due to its cost-effectiveness and non-invasiveness. 

Numerous recent research provide a range of options for AD detection using MRI images by 
utilizing machine learning techniques, especially deep learning methodologies. 
The accuracy of identifying various forms of dementia is improved when machine learning 
(ML) is used to neuroimaging. Certain pre-processing steps must be taken in order to 
facilitate the use of ML algorithms. These steps include using a classifier method, 
reducing the dimensionality of features, and extracting and choosing features.
Each of these stages is essential to the ML classification process \citep{ieracitano2020novel}. 
Convolutional Neural Networks (CNNs), which can automatically extract a variety of relevant semantic 
information from pictures, have been progressively encompassed to MRI diagnostic tasks due 
to the quick advancement of computing power and computer vision. MRI is regarded the gold standard, 
due to its limited ionizing radiation, good tissue contrast, and great spatial resolution \citep{moser2009magnetic}. 
Developing a trustworthy computer-aided diagnostic system that can analyze MRI images and detect AD is essential. 
As such, any AI algorithm used to detect dementia in an MRI scan must be able 
to classify the stage of the disease so the person can get the proper treatment. 
Furthermore, many AI models, such as neural networks, are 'black boxes', which means the 
researcher is unsure exactly how the model makes the classification based on 
input \citep{ebrahimi2021convolutional, salehi2020cnn}. In the case of an MRI-based 
diagnosis, the developer is unable to show what regions of the scan the model is using 
to make its classification. This can cause problems in the diagnosis process, as the medical 
professional is unable to verify the diagnosis or see the logic behind the AI’s 
classification. To solve this problem, the discipline of Explainable Artificial Intelligence 
(XAI) was created, which creates new kinds of interpretable models or provides explanations 
for the predictions made by black-box systems. In the imaging field, creating a unique 
heatmap for every input image and highlighting the significance of each pixel in the 
ultimate classification choice is a popular method. Such a technique can be an efficient 
tool that can be effortlessly incorporated into prospective computer-aided diagnosis 
software to offer human-intuitive justifications for the classifier's choices \cite{bohle2019layer}.

Despite the severity of these obstacles, scientists and medical professionals are 
making every effort to overcome them in the context of classifying neurodegenerative diseases. 
The main focus of these initiatives is the use of AI methods that help produce precise diagnostic results. 
This research has focused on those those problem and attempted 
to work through it. The major contributions of this research are as follows:
\begin{itemize}
\item We present a novel deep learning architecture that employs an advanced convolutional neural Network (CNN), with multiresidual block, custom spatial attention block,  
grouped query attention and Multihead attention.
\item Four datasets were used in the development and testing of a generalized model, 
which led to lower computing costs and parameters. The model performed exceptionally well 
in the diagnosis of AD in spite of these optimizations.
\item Rather than working on a single plane (axial), our model was trained and evaluated on other two planes i.e. sagittal, coronal, and also combining all of the planes.
\item We utilized five state of the art XAI techniques such as, GradCAM, ScoreCAM, Faster ScoreCAM, and  XGradCAM to provide
interpretable explanation for predictions generated by our proposed approach. Because it allowed 
us to confirm the significance of particular features, compare and contrast the data acquired 
from other methodologies, and highlight the clinical relevance of our model, 
the interpretability was compared with other current approaches.
\item We have trained and evaluated our model in different class scenarios such as 5 stages of AD, 4 stages, 3 stages, and binary classification of AD. 
\item Our results show that, in spite of the dataset's class imbalance, our model 
performed exceptionally well thanks to efficient feature engineering and modeling techniques. 
This could have important ramifications for enhancing ML applications and promoting their 
use in the clinical setting.
\end{itemize}
The remaining paper is organized as follows: \autoref{works} presents a comprehensive analysis of the current literature on the classification of AD. 
\autoref{methodologies} discusses the detail overview of dataset, their pre-processing techniques and  brief overview of proposed architecture with its parameters 
and implementation details followed by the explanation of the result in \autoref{experimental_results} with explainability analysis. Finally \autoref{conclusion} 
gives the conclusion with future direction.

\section{Related Works}\label{works}
The approach of using machine learning to AD progression modeling is intricate. Most AD studies relies on MRI analysis.
\cite{kavitha2022early} incorporated voting classesfier with different ML models on OASIS 
dataset namely: Decision Tree (DT), Random Forest (RF), Support Vector Machine (SVM), and 
Extreme Gradient Boosting (XGBoost). To identify most relevant features from the dataset, 
some feature selection (FS) algorithms was used. Using the same dataset, another work was 
done by comparing different ML models, and it was observed that the SVM model gains highest 
accuracy amongst other models \citep{bari2021comparative}. 
Another work incorporated PET data along with MRI scan data to predict the progression from 
MCI to AD using state-of-the-art ML models (SVMRF, Gradient Boosting (GB)) 
\citep{beltran2020inexpensive}. \cite{rallabandi2020automatic} implemented nonlinear SVM 
on whole brain MRI scan to classify the progression of brain  stages of impairment from 
cognitive normal (CN) to AD. Similarly, in another work by \cite{shahbaz2019classification}, 
experiment was conducted on ADNI datasets using different ML models like Naive Bayes(NB), 
DT, rule reduction, and generalized linear model. \cite{almohimeed2023explainable} suggested 
a unique multi-level stacking approach to predict various AD phases by combining 
heterogeneous ML models. Six base model was used in this study namely- RF, DT, Logistic Regression (LR), 
k-nearest Neighbors(KNN), and NB. Model interpretability was also introduced to ensure the 
efficiency, effectiveness of proposed model using Explainable Artificial intelligence(XAI). 
Similarly, data-level fusion using Alzheimer’s Disease Research Center (ADRC) clinical data, 
brain Magnetic Resonance Imaging (MRI) segmentation data, and psychological assessments data 
has been used in the work of \cite{jahan2023explainable}. Different ML models was used along 
with SHAPELY for model interpretation.  Fusion of different modalities data could provide a 
better view on patient's situation and could be able to enhance the accuracy of progression 
model. \cite{zhang2012multi} suggested a generic technique known as multimodal multitask (M3T) 
learning that combines multimodal data (such as MRI, PET, and CSF) to predict various medical 
scores (such as the MMSE, ADAS, and diagnostic feature) at the same time. 
They used baseline data from a limited number of modalities and SVMs.
\cite{el2021multilayer} developed a two-layered explainable machine 
learning model for the categorization of AD. Data from eleven different modalities—genetics, 
medical history, MRI, Positron Emission Tomography (PET), neuropsychological battery, 
cognitive scores, etc.—were combined in this multimodal method. Here, the Random Forest (RF) 
classifier was employed in the first layer for multiclass classification, and the SHAP 
framework was utilized to interpret the findings. Binary categorization, which likely 
classified MCI to AD, occurred in the second layer. 
The majority of AD research only looks at a small number of characteristics, which may 
not be enough to fully comprehend this complicated illness \citep{ding2018hybrid}. Though these works show very 
good performance , but these technique solely are dependent on good extracted features, 
which is time consuming and needs experts' opinions. Deep learning (DL) can improve this 
challenging modeling task's performance \citep{zhang2021survey}. 

Most of the state-of-the-art DL models either utilize pre-training or non-pretraining methods. 
Some literatures have used transfer learning techniques on models including VGG16, ResNet, 
AlexNet, GoogleNet, and etc. \cite{jain2019convolutional} proposed a CNN transfer learning 
architecture (VGG-16) for AD diagnosis. The results of their experiment are based on a three-way 
categorization in the ADNI database. In another work, \cite{lee2019using} introduced a deep CNN data permutation 
strategy that uses resting-state MRI to classify AD. To make best use of AlexNet's advantages, they suggested 
slice selection. Their data permutation approach enhanced the overall classification accuracies for AD 
classification, according to their experimental results. In another work, rather than using conventional thresholding, 
\cite{abdulazeem2021cnn} utilized adaptive thresholding that changes dynamically over 
the image. It helps to adapt to changes in lighting conditions of images. Simple CNN 
architecture has been used for predicting AD progression using ADNI dataset 
and several experiments was done by changing parameters to get the best result. 
\cite{shamrat2023alzheimernet} worked with some basline pre-trained architectures such as 
ResNet50, MobileNetV2, VGG16, AlexNet and Inception-V3 by fine-tuning the parameters and 
finally proposed a modified InceptionV3 as the baseline outperforms other architectures. 
Two different CNN architecture was proposed by \cite{el2024novel}consisting different 
filters for different feature extraction, which are then concatenated before the decision layer. 
Thus, both knowledge of task specific features is enabled by complimenting the models to each other. 
Adaptive Synthetic Sampling Approach (ADASYN) has been utilized for solving the data imbalance problem.
\cite{arafa2024deep} proposed a simple CNN architecture along with another 
method that uses a pre-trained VGG16 architecture and compared the output of both on different parameters 
by applying some pre-processing on dataset and data augmentation to mitigate the class imbalances.  
Similarly, in another study, a joint learning technique that combines both deep learning and machine learning approach was proposed by \cite{abuhmed2021robust}. Two hybrid architecture was introduced that is able 
to predict the AD progression even after 2.5 years later from the multimodal time series data. The two methods are: deep feature-based
learning which utilizes multivariate BiLSTM architecture for feature learning following the ML models for prediction and multitask 
regression-based learning that first learns the seven regression tasks and then ML models for prediction. \cite{lee2019predicting} 
employed a multimodal recurrent neural network (RNN) model for predicting the progression of AD from the stage of 
MCI. In this method, the authors combined cross-sectional neuroimaging data, demographic data, and the participants' 
longitudinal cerebrospinal spinal fluid (CSF) and cognitive performance biomarkers. In another study, multiclass 
classification was performed on volumetric 
18F-FDG PET images collected from ADNI dataset \citep{de2023explainable}. Saliency Map (SM) and Layerwise Relevance 
Propagation (LRP) are two distinct post hoc explanation strategies that were used to test the performance of a 3D 
CNN architecture.Similar to this, a modified Resnet18 deep learning architecture is trained concurrently on the two 
datasets using a unique heuristic early feature fusion approach that concatenates PET and MRI images. 
\cite{ansingkar2022efficient} adopted a hybrid equilibrium approach to improve a capsule encoder network for AD diagnosis.
A hybrid CNN model, for feature extraction, followed by a 
KNN with Bayesian optimization for thewas incorporated in another study \citep{lahmiri2023integrating}. 
The validation dataset's sample size is constrained, though. In addition to the multi-layer CNN structure, 
autoencoders were developed in this study to increase accuracy by utilizing fine-grained abstract information.

To increase the efficacy of CNN-based models, enhanced attention-guided multiscale features can be used as the attention mechanism canprovide 
more focused features on weights and provide proper interpretability. In several 
literature, different types of attention mechanism such as channel-wise squeeze  and excitation module, spatial attention, self attention, 
and different task specific attention module has been used \citep{liu2022diagnosis, zhang2021explainable, zhang20213d, jin2019attention}. 
These studies shows the effectiveness of attention modules in sense of model performance and interoperability. By addressing other 
literatures lackings on attention to high or low-level spatial features,  \cite{tripathy2024alzheimer} proposed a 
modified spatial attention guided network with the depth separable convolutional layer. The proposed approach 
introduces improved spatial attention block to enhance 
the spatial attention maps by combining multiple feature cues. These maps are used to extract multiscale, spatially directed features. 
The maps goes through the separable CNN layer with skip connection and multilayer spatial 
attention feature is combined forming scale invariant features that goes into the classification stage. 
Most of the proposed attentionmethods work with single spatial scale which are unable to learn a discriminative 
feature representation of AD patients. So, a multi-scale information integration block has been proposed 
by \cite{wu2022attention} using dilated convolution and soft attention mechanism. It helps to capture better 
multi-scale features with fewer computations.  3D MRI images consists better multi-sclare features, which was studied also in several research.    
A powerful interpretable network naming TabNet, which was built upon transformer model, was proposed for AD classification using 3D 
t1-weighted brain MRI dataset \citep{park2023development}. The dataset was processed to extract 104 brain subregion and 68 cortex region 
by the help of VUNO Med-deepbrain , which is a deep learning based software to analyze brain parcellation and quantification. 
\cite{kang2023interpretable} proposed a framework for extracting high dimensional brain ROI's with the help of CNN. The DL based biomarkers 
has been used by Explainable boosting machines (EBM), a tree-based Generalized Additive Model (GAM), to predict AD progression. As baseline 
methods to extract biomarkers from brain, Glo-CNN and Loc-CNN has been incorporated. 
In another study, \cite{dhaygude2024knowledge} proposed a deep 3D convolutional network incorporating multitask learning and attention 
mechanism. Two auxiliary subtasks such as Clinical Dementia Rating (CDR) scale score regression and Mini‐Mental State Examination (MMSE) 
score regression is added to optimize the accuracy of AD classification. For the purpose of randomizing clinical trials for AD and assessing 
the effect of reducing allocation bias on trial efficiency, a second CNN-based model with a self-attention mechanism was 
presented by \cite{wang2024multimodal}. Non-image data, including as demographics, cognitive test results, and 
biomarkers, were added to T1-weighted pictures in order to forecast changes in the patient's health. 
In the study by \cite{mahim2024unlocking} for AD classification vision transformer (ViT) with the combination of GRU has been incorporated. 
The ViT extracts crucial features from the images and GRU creates strong correlation between those features giving the architecture a 
push to higher performance. Different XAI techniques was experimented namely LIME, SHAP and attention maps for clear model interpretation. 
\cite{ahanger2024alzhinet} introduced Alzhinet that uses self-attention mechanism on 3D volumetric MRI data to predict AD stages. 
The baseline pretrained VGG16 model was utilized rather than more complex structures like ResNet59 and Inception model. The three 
view of brain namely axial, sagittal and coronal was processed with several steps like skull stripping, denoising, normalization, and 
registration nad used for classification. Before entering the VGG16 network as a 3-channel, 224x224 pixel 2D picture, each MRI slice of the scan is scaled along the axial plane's z-axis.
\cite{adarsh2024multimodal} presented a CNN framework and a within-class-similar discriminative dictionary learning 
technique that uses anatomical and structural similarities between similar pictures in the training set to minimize 
misclassification. To verify and enhance the classification's accuracy, a transfer learning procedure and a decision 
tree mechanism are employed. To create a model that is comprehensible and whose judgments can be relied upon, 
LIME and CAM are utilized.
\section{Materials and Methods} \label{methodologies}
\begin{figure}[ht]
  
  \centering
  \graphicspath{./fig/method.png}
  \includegraphics[width=0.8\textwidth, height=0.25\textheight]{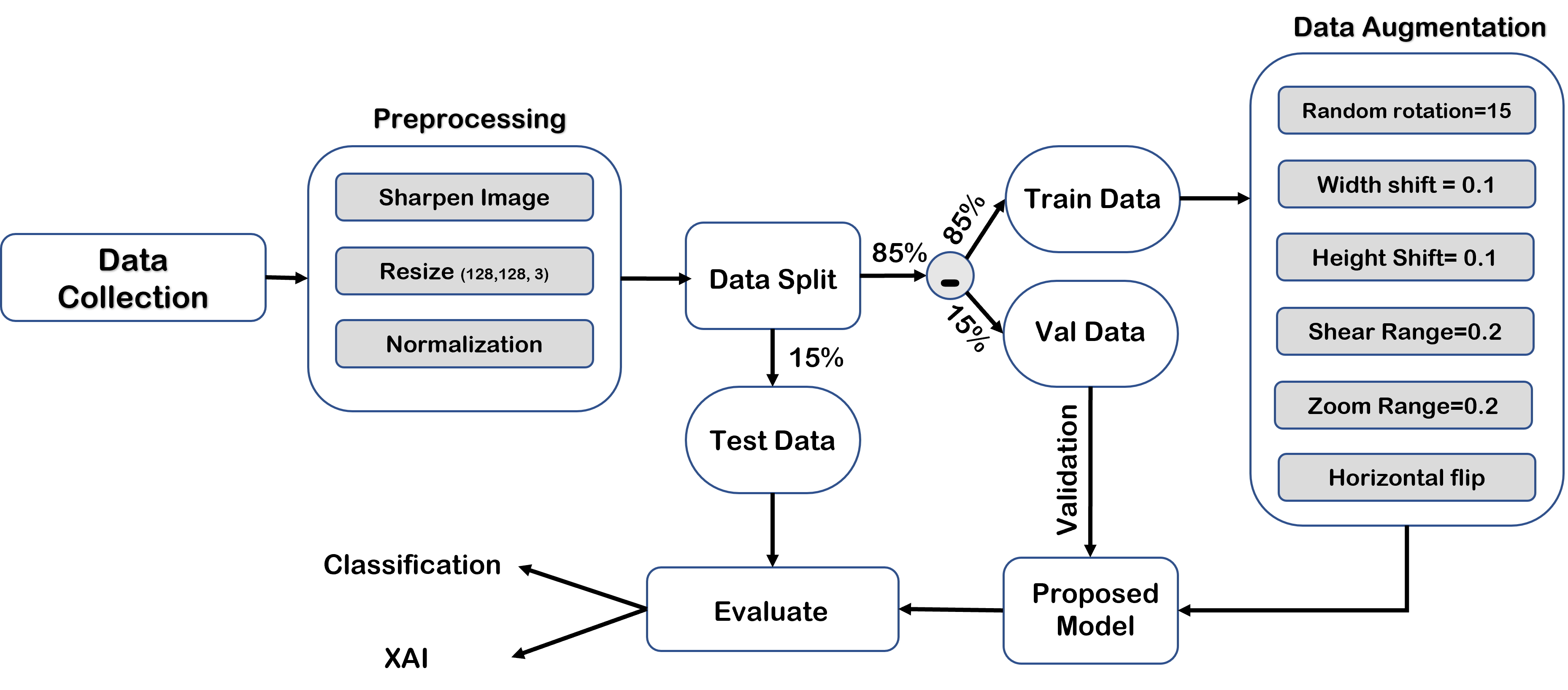}
  
  \caption{An overview of the proposed method}\label{methodology}
  
\end{figure}

We introduce a novel framework intended to accomplish two important objectives in the categorization of medical images: interpretability and high accuracy. 
For efficient classification, our system combines advanced CNN techniques, kernel approaches, and explanation algorithms. Each of the framework's several 
parts addresses a distinct set of difficulties in the categorization and interpretation of medical images. We have carefully processed medical photos before feeding 
them into the neural network. Data is splitted in three protions: train, validation, and test. First the test data, which is 15\% of the total data, is separated. 
Then, from the remaining data train and validation data has been splitted in 85:15 ratio. Image augmentation is performed only on the training data to mitigate the 
problem of huge class imbalancing. A specially created CNN that is trained for categorization, sits at the heart of the system. We have included a Group Query Attention layer, 
a multihead attention layer, and a bespoke Spatial Attention layer to improve the procedure. The network can also extract characteristics at different degrees of 
detail because of the integration of a multi-residual structure. The proposed network is trained and evaluated on train and validation data respectively, and later tested on the test dataset
in different sets of experimentation. The overall overview of the proposed methodology is shown in \autoref{methodology}. 
\subsection{Dataset}
\subsubsection{Kaggle Dataset}
\begin{figure}[ht]
  
  \centering
  \graphicspath{./fig/fig_2.png}
  \includegraphics[width=0.8\textwidth, height=0.15\textheight]{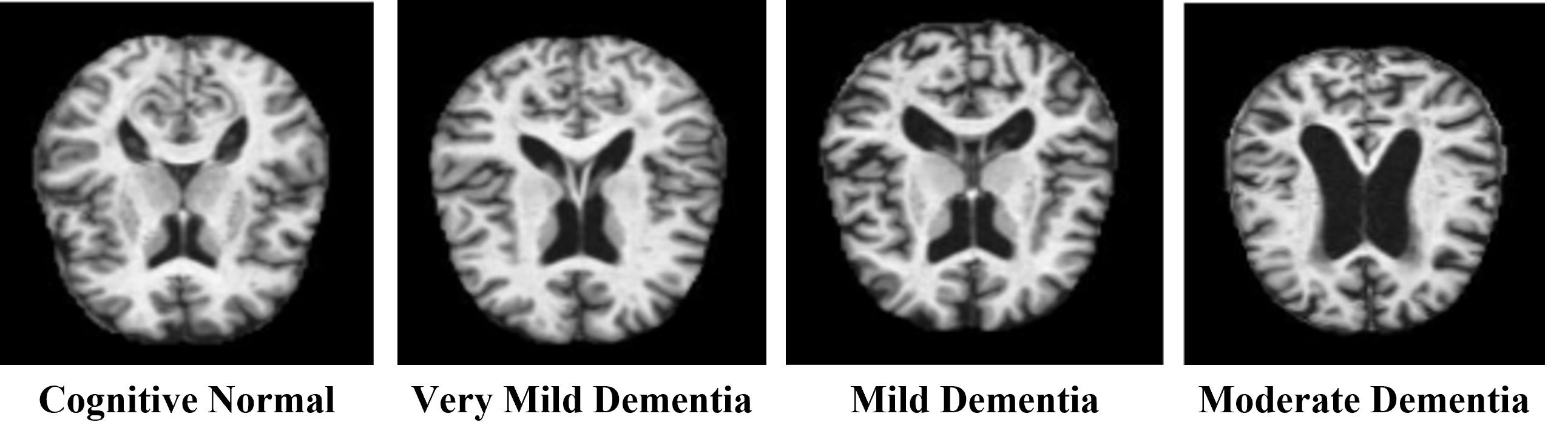}
  
  \caption{Sample images of kaggle datasets on 4 classes}\label{sample_image}
  
\end{figure}

This research uses an publicly available dataset to detect the progression of AD from the open-source data science platform Kaggle \citep{kaggle_data}. 
The dataset comprises a total of 6,400 images, outsourced from different hospitals, websites, and public repositories. The data set consists
 of four classes, namely moderate demented, mild demented, very mild demented,  and non-demented each having a subjects of 2, 28, 70, and 100, respectively. 
 From each subject scan 32 slice images are extracted. Some sample images of four classes of this dataset are shown in \autoref{sample_image}.
\subsubsection{Open Access Series of Imaging Studies (OASIS)}
The OASIS dataset hosted by Neuroimaging Informatics Tools and Resources Clearinghouse-Image Repository (NITRC-IR), 
provides community support for easy access. In this experiemnt, we used the OASIS-1 dataset. The dataset contains a 
cross-sectional collection of 416 subject from middle aged person to elderly person (18 - 96 years old). For each subject, 3 or 4 individual T1-weighted 
MRI scans obtained in single scan. The subjects include both men and women, additionally it was kept in mind that all the subjects are right handed.
A clinical diagnosis of very mild to severe AD has been made for 100 of the participants over 
60. Furthermore, 20 nondemented participants who were photographed on a follow-up visit within 90 days of their 
original session make up the reliability data set. From this dataset used in this study, each subject's scans is sliced 
along sessions are included. The z-axis 
or plane into 256 pieces, and slices ranging from 100 to 160 are selected from each recordings. Totalling the dataset of 80,000 images.

\subsubsection{ADNI-1}
The Alzheimer's Disease Neuroimaging Initiative (ADNI) is a large, collaborative effort established to help researchers track the progression of 
AD using various biomarkers, including neuroimaging, genetic markers, and cognitive measures. ADNI-1 specifically focused on collecting baseline data 
and longitudinal follow-ups of participants diagnosed with AD, MCI, and cognitively normal (CN) subjects One of the key imaging techniques used in ADNI-1 is MRI, 
performed with both 1.5T scanners. This dataset contains recording of 643 subjects. The scans of each subject was divided into three planes  namely axial, coronal 
and sagittal. From each plane among different slices only middle 20 slices were converted into image \citep{petersen2010alzheimer}. In the \autoref{axial_images} below, 
shows a brain slice of three plane.
\begin{figure}[ht]
  
  \centering
  \graphicspath{./fig/fig_2.png}
  \includegraphics[width=0.8\textwidth, height=0.25\textheight]{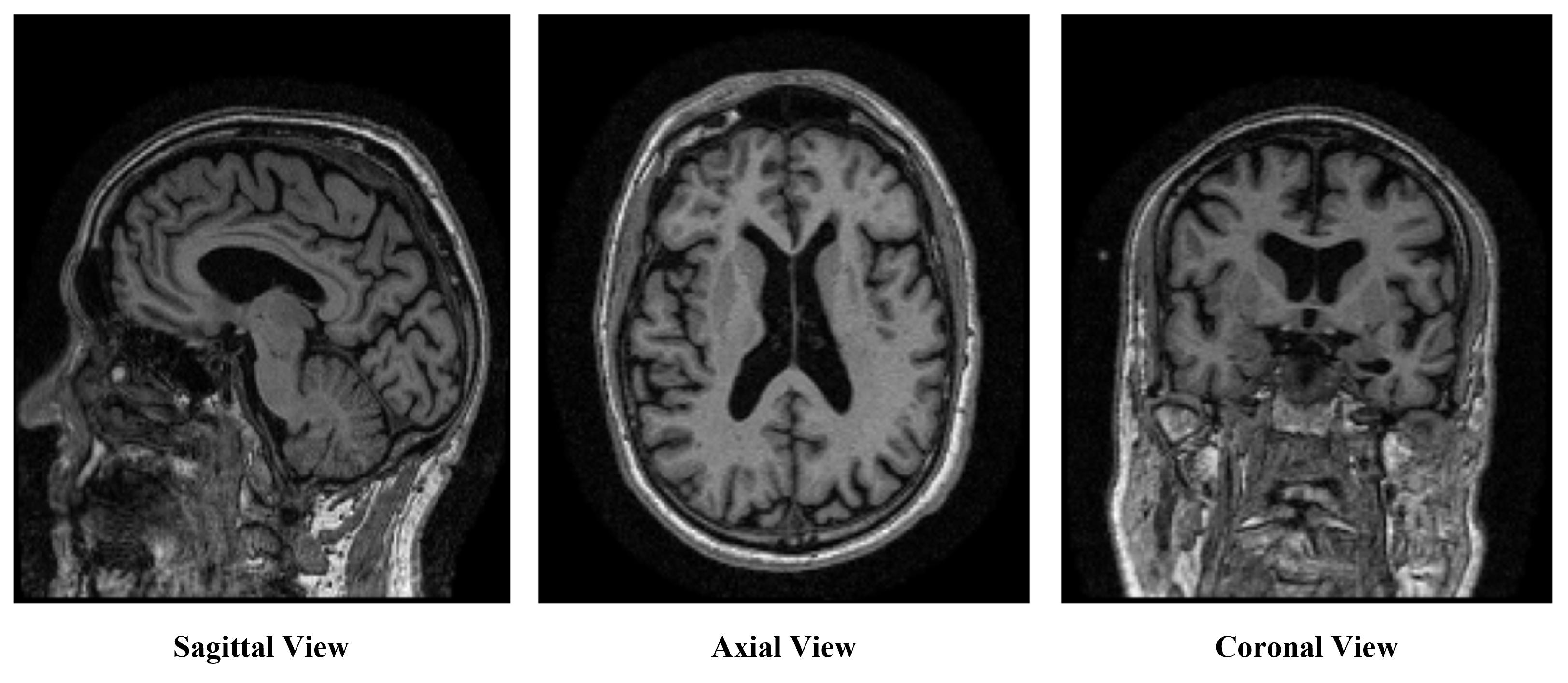}
  
  \caption{Sliced image of three plane of brain }\label{axial_images}
  
\end{figure}

\subsubsection{ADNIGO/ADNI-2}
ADNI-2, also known as ADNIGO (ADNI Grand Opportunity), expanded the original ADNI-1 study to further investigate the progression of 
AD and improve early detection of the disease. ADNI-2, which ran from 2010 to 2016, introduced several advancements in imaging technology and included a 
wider array of participants, including individuals with early mild cognitive impairment (EMCI) and late MCI (LMCI), to capture an even broader spectrum of the cognitive 
aging process \citep{petersen2010alzheimer}. This dataset contains 5 classes of T1 weighted and T2 weighted processed datasets. Only axial plane has been selected from the scans of each 
subject. Among the whole slices of the axis, middle 16 slices has been considered. 

\begin{table}[t]
\caption{Overview of sample image distribution of each class in datasets}
\centering

\scriptsize\setlength\tabcolsep{8pt}
\begin{tabular}{l c c c c c c  }
\hline
\textbf{Classes}& \multicolumn{6}{c}{\textbf{Datasets}}\\
 & \textit{Kaggle} & \textit{OASIS}& \multicolumn{3}{c}{\textit{ADNI-1}}& \textit{ADNI-2}  \\ \hline
 & & & Axial& Sagittal& Coronal& \\ \hline
 \hline
 \makecell[l]{Cognitive Normal\\ (Non-Demented)} &3200 &67,222 & 3880& 3880& 3880&8,650 \\ 
 \makecell[l]{Early Mild Cognitive Impairment\\ (Very Mild demented)} &2240 &13,725 &- &- &- & 480\\
\makecell[l]{Mild Cognitive Impairment\\ (Mild Demented)} & 896&5002 &6320 & 6320& 6320 & 1155\\ 
 Late Mild Cognitive Impairment &- &- & -& -& - & 144\\
 \makecell[l]{Alzheimer Disease\\(Moderate Demented)} & 64&488 & 2660& 2660& 2660& 8346\\ \hline
 \hline
 \end{tabular}\label{class_numbers}
 \end{table}
\subsection{Data partition and class distribution}
In our experiment, we evaluate four open-access datasets with a variety of class segments. For each study, we ensure the test data remains 
separate and consistent across all experiments for each dataset. To achieve this, we first split the data into 85\% and 15\%, reserving the latter 
for testing. We then further divide the initial 85\% into training and validation sets with same 85:15 ratio. By using a consistent random state of 43 during the first stage of 
splitting, we ensure the train, validation, and test data remains the same for every dataset across all experiments.

Each dataset consists of different classes. We experiment on various extents to evaluate our proposed method. For both the OASIS and Kaggle datasets, four classes 
are available: Moderate Demented, Non-Demented, Very Mild Demented, and Mild Demented. We experiment with both multiclass and binary classification. In the 
four-class setup, the classes remain unchanged. For the three-class setup, we combine Mild Demented and Moderate Demented patient data. For binary classification,
 we merge Mild Demented and Very Mild Demented samples with Moderate Demented samples. In the ADNI-GO dataset, there are five classes: AD, CN, EMCI, MCI, and LMCI. 
 We conduct two multiclass experiments using five classes and three classes. For the three-class setup, we combine EMCI and LMCI data with MCI. In the ADNI-I dataset, 
 we create four datasets using each brain axis and one with all axis. Using this dataset, we conduct only multiclass classification using the existing three 
 classes: AD, MCI, and CN.

\subsection{Preprocessing and Data Augmentation}
Data preprocessing is a vital step in image classification tasks, ensuring the data is properly prepared before putting inside the model. Our study utilizes multiple datasets, 
including the ADNI-I and ADNI-GO datasets, which consist of DICOM format brain scans. These scans are 3D images with dimensions of  (256×256×256), captured in axial, 
coronal, and sagittal planes. However, not all slices are relevant for analysis. From the ADNI-I dataset, we selected 20 middle slices, while for the ADNI-GO dataset, 
16 middle slices were extracted. These slices were then converted into JPEG format for further processing. Subsequent preprocessing steps were consistent across all datasets. 
First, the images were converted to RGB format, followed by rescaling and normalize pixel values between 0 and 1. This normalization is essential, as neural networks 
are sensitive to input data scales, and rescaling helps the model effectively learn patterns. Additionally, all images were resized to a standardized dimension of (128×128x3) to 
ensure the uniformity across the dataset and improve processing efficiency. We use the sharpening filter to enhance the edges and fine details. The central pixel is emphasized 
with a value of five, while neighboring pixels are subtracted using negative weights (-1). This results in a stronger contrast at the edges, making them more defined and 
more sharpened across the overall image. The overall preprocessing stage can be seen in \autoref{methodology}.

The data augmentation process plays a critical role in addressing class imbalancing and enhancing the diversity of the dataset, which ensures a more robust model. 
The augmentation parameters, in this experiment, are carefully selected to promote variability.
We used a rotation range of 15 degrees, width and height shift ranges of 0.1 (10\% of the image dimensions), a 
shear range of 0.2, and a zoom range of 0.2 (up to 20\%).
We also apply random horizontal flips to further diversify the images. To do this, Tensorflow's built-in TrainDatagen function is used. 
For the multiclass classification task in this experiment, a strategic augmentation strategy was adopted to balance the dataset. 
In particular, the underrepresented class's sample size is raised to equal that of the second-largest class. The 
underrepresented class receives one-third as many samples as the second-largest class when there is a very significant difference in class numbers. 
This targeted approach ensures that the model is trained on a more balanced and representative dataset, thereby improving its ability to learn more generalized 
patterns across all classes. It is important to note that the augmentation is done after the data split and only on training dataset. The validation data and 
test data are free from any data augmentaion.

\subsection{Proposed Architecture}
\begin{figure}[H]
  
  \centering
  \graphicspath{./fig/blocks.png}
  \includegraphics[width=0.9\textwidth, height=0.6\textheight]{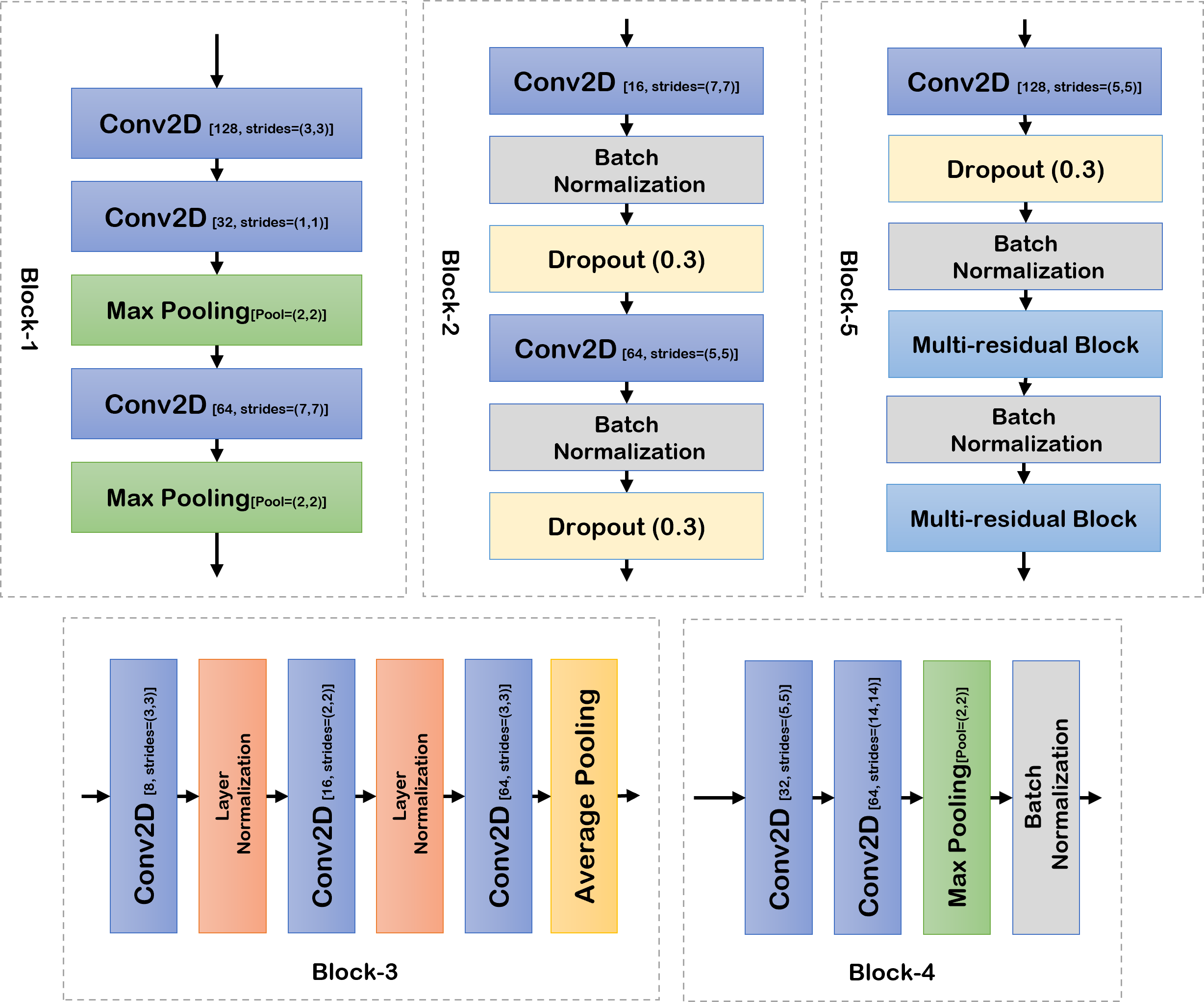}
  
  \caption{Structure of diffferent convolutional blocks' structure}\label{blocks}
  
\end{figure}

\begin{figure}[H]
  
  \centering
  \graphicspath{./fig/model.png}
  \includegraphics[width=\textwidth, height=0.7\textheight]{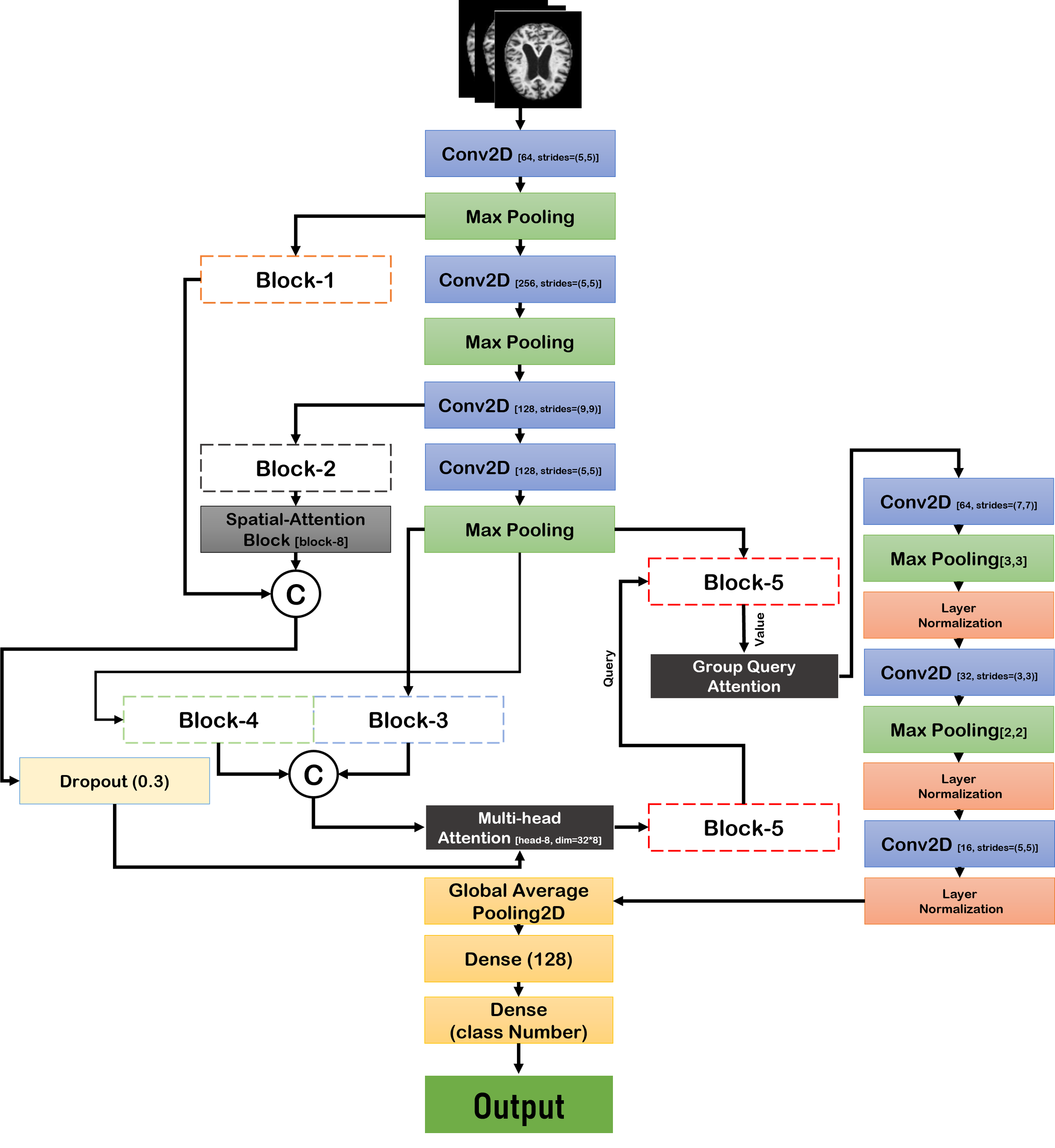}
  
  \caption{Proposed Model Architecture}\label{proposed_model}
  
\end{figure}

The proposed architecture combines multiple advanced components to achieve high accuracy and interpretability in alzheimer progression  classification. It begins with 
processed input images  through a series of Conv2D layers with varying filter sizes and strides, followed by Max Pooling for feature extraction and dimensionality reduction. 
There are several convolutional layer block in the structure, consisting of different orientation of layers and normalization to capture the best porssible features from the input. 
\begin{figure}[h]
  
  \centering
  \graphicspath{./fig/norms.png}
  \includegraphics[width=0.8\textwidth, height=0.3\textheight]{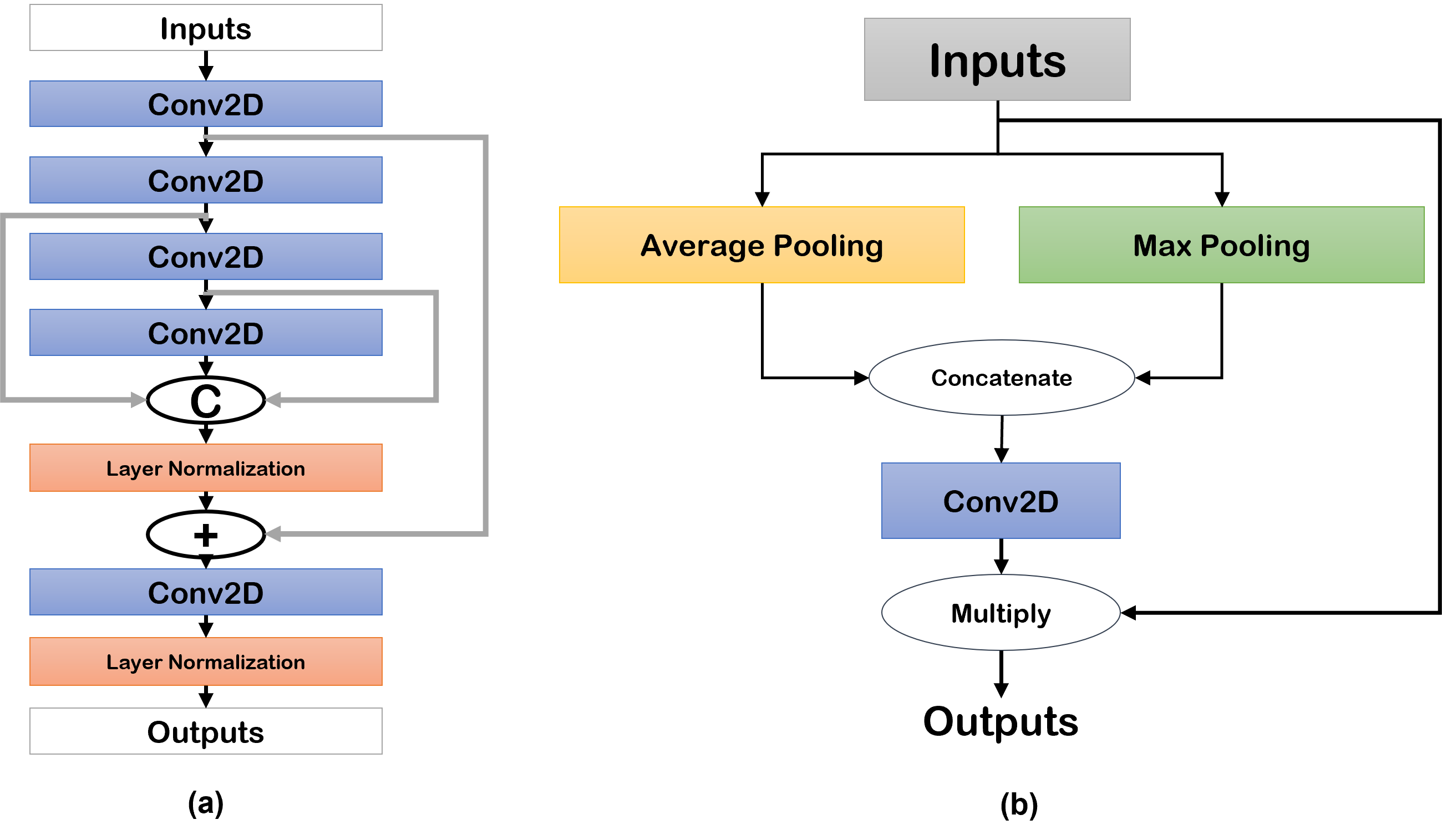}
  
  \caption{Architecture illustration of - (a) Multi-residual block, and (b) Custom spatial Attention block}\label{norms}
  
\end{figure}
A custom Spatial Attention block enhances focus on critical regions, while a Multi-Residual structure ensures effective feature extraction and learning. 
The Group Query Attention layer refines feature extraction by modeling inter-feature relationships, aided by Layer Normalization for stability. 
A Multi-Head Attention layer captures long-range dependencies, improving feature representation. To prevent overfitting, a Dropout layer is included, and 
Global Average Pooling reduces feature dimensions into compact vectors. Finally, Dense layers process the extracted features and with the last dense layer performs the classification. 
This design effectively integrates attention mechanisms, normalization, and pooling strategies to handle complex medical imaging tasks with precision and clarity.The proposed 
architecture and its internal components are illustrated in Figures \ref{blocks}, \ref{norms}, and \ref{proposed_model}. 

\subsection{Hyperparameters and Optimization Strategy}

As we optimize the proposed model, our focus gradullay shifts to hyperparameter tuning, a crucial step in achieving optimal performance. This process involves identifying 
hyperparameter values that closelyyields the best possible results.We have uniformed the hyperparameters and 
initialization processes to provide uniform and equitable assessment across various neural network topologies.
In all experiments, the same hyperparameters are applied uniformly. 
\autoref{parameters} lists the hyperparameters that we have used in experiments along with their corresponding values.
\begin{table}[H]
\caption{Hyperparameters of our proposed method and its values}
\centering

\scriptsize\setlength\tabcolsep{8pt}
\begin{tabular}{l c||l c }
\hline
\textbf{\textit{Hyperparameters}} & \textbf{\textit{Values}} & \textbf{\textit{Hyperparameters}} & \textbf{\textit{Values}}\\
 \hline
 Epochs & 50 &\makecell[l]{Initial Learning\\ Rate} & 0.0001\\ 
 
 Batch Size & 8 &  Factor & 0.7\\

 Image Size & 128 x 128 x 3 &patience & 7\\

 Callback & ReduceOnPlateau & min\_lr & \ensuremath{1e-6}\\
 Loss & Categorical cross-entropy &Optimizer& Adam \\\\
\hline

 \end{tabular}\label{parameters}
 \end{table}

\subsection{Evaluation Procedures and Training}
The performance of the proposed model is assessed using various metrics, including accuracy, precision, recall, F1-score, and sensitivity. These metrics are derived from 
the confusion matrix, which provides values such as True Positives (TP), True Negatives (TN), False Positives (FP), and False Negatives (FN) to reflect the model's 
performance. Additionally, the Area Under the ROC Curve (AUC-ROC) is calculated using the True Positive Rate (TPR) and False Positive Rate (FPR) to further validate the 
model. Root Mean Squared Error (RMSE) is also employed as an evaluation metric.
\begin{equation}
  \text{Accuracy} = \frac{TP + TN}{TP + TN + FP + FN}
  \end{equation}
  \begin{equation}
  \text{Precision} = \frac{TP}{TP + FP}
  \end{equation}
  \begin{equation}
  \text{Recall} = \frac{TP}{TP + FN}
  \end{equation}
  \begin{equation}
  \text{F1-Score} = \frac{2 \cdot \text{Precision} \cdot \text{Recall}}{\text{Precision} + \text{Recall}}
  \end{equation}
  \begin{equation}
  \text{Sensitivity} = \frac{\text{True Positives (TP)}}{\text{True Positives (TP)} + \text{False Negatives (FN)}}
  \end{equation}
  \begin{equation}
  \text{TPR} = \frac{TP}{TP + FN}
  \end{equation}
  \begin{equation}
  \text{FPR} = \frac{FP}{FP + TN}
  \end{equation}
  \begin{equation}
  \text{AUC} = \int_0^1 \text{TPR} \, d(\text{FPR})
  \end{equation}
  \begin{equation}
  \text{RMSE} = \sqrt{\frac{1}{n} \sum_{i=1}^n (y_i - \hat{y}_i)^2}
  \end{equation}
  where \ensuremath{y_i} is the actual value and \ensuremath{\hat{y}_i} is the predicted value. The less the RMSE, the better is the performance of the architecture.

The proposed network is implemented in TensorFlow 2.0 and Keras, using Python 3.11. The experiments are conducted on a system equipped with an NVIDIA RTX 3060 GPU, 
an Intel Core i7 (3.5 GHz) CPU, and 128 GB of RAM.

\subsection{Explainable Artificial Intelligence}
To aid in the diagnostic process, an explainable AI (XAI) system was developed to identify the regions of the MRI scan that the CNN is using to classify the image. 
It gets harder to comprehend the logic underlying AI models' judgments as they get more complicated and smart. This raises possible problems in the medical field, 
as AI systems have the ability to make judgments that might change people's lives. XAI is very important in the context of medical picture categorization \citep{adarsh2024multimodal}. 
XAI can result in more accurate assessments about patient care and treatment plans by assisting medical 
practitioners in comprehending and interpreting the decisions made by AI systems. 
It is also crucial for ethical and regulatory grounds, as medical organizations and authorities frequently demand 
openness in the way AI systems make decisions in medical applications to guarantee adherence to rules and norms. 
Several explainable AI techniques, such as LIME, SHAP, and 
attention map were applied to improve the interpretability and transparency of DL models used for AD detection. In this experiment, we used Gradient 
Class Activation Mapping (GradCAM), Score-CAM, Faster Score-CAM, and XGRADCAM. To get the output for the GradCAM, the feature 
importance with respect to each feature map \( C_f \) is obtained as:
\begin{equation}
W_f = \sum_{x} \sum_{y} \frac{\partial S_c}{\partial C_f}
\end{equation}
The importance weight is calculated as:
\begin{equation}
H_f = \text{ReLU} \left( \sum_{y} \left( \frac{1}{N} W_f \right) \times C_f \right)
\end{equation}
The score for class \( c \) is given by:
\begin{equation}
S_c = \sum_{f} \frac{1}{z} \sum_{x} \sum_{y} C_f
\end{equation}
The output of the global average pooling is defined as:
\begin{equation}
G_f = \frac{1}{N} \sum_{x} \sum_{y} C_f
\end{equation}
Rewriting the score for class \( c \):
\begin{equation}
S_c = \sum_{f} W_f G_f
\end{equation}
The gradients for the class score with respect to the feature maps are computed as:
\begin{equation}
\frac{\partial G_f}{\partial C_f} = \frac{1}{N}
\end{equation}
Finally, summing both sides of the equation we get the output of GradCAM:
\begin{equation}
\frac{\partial S_c}{\partial G_f} = W_f \frac{\partial S_c}{\partial C_f} N
\end{equation}
The simplified equation will be: 
\begin{equation}
S_C = \text{ReLU} \left( \sum_{k=1}^{K} \alpha_k \cdot A_k \right)
\end{equation}
where:
\begin{equation}
\alpha_k = \frac{1}{Z} \sum_{i,j} \frac{\partial y_c}{\partial A_k(i,j)}
\end{equation}
The Score-CAM equation is written as:
\begin{equation}
S_c = \text{ReLU} \left( \sum_{k=1}^{K} \omega_k \cdot A_k \right)
\end{equation}
Where the weight \( \omega_k \) is calculated by:
\begin{equation}
\omega_k = \frac{\exp(f_k)}{\sum_{k'} \exp(f_{k'})}
\end{equation}
The XGrad-CAM equation is written as:
\begin{equation}
S_c = \text{ReLU} \left( \sum_{k=1}^{K} \alpha_k^{X} \cdot A_k \right)
\end{equation}
Where the weight \( \alpha_k^{X} \) is given by:
\begin{equation}
\alpha_k^{X} = \frac{1}{Z} \sum_{i,j} \left( \frac{\partial y_c}{\partial A_k(i,j)} + \eta \right)
\end{equation}

\section{Experimentation and Analysis} \label{experimental_results}
\subsection{Comparative Analysis}
For our analysis, we use four datasets, each containing different sets of samples for various classes. To evaluate the performance of our model, we use several metrics: 
accuracy, precision, recall, F1-score, RMSE, AUC-score, and sensitivity. In our first experiment, we use the Kaggle dataset, which contains 6,400 brain image samples. 
Originally, it has four classes. We experiment on two multiclass classifications and one binary classification. For 4-class classification, we outperform existing approaches
 with an accuracy of 99.66\%, an RMSE score of only 0.0547, and a sensitivity of 99.21\%. For 3-class classification, we achieve an accuracy of 99.63\%. Our proposed model 
 achieves a perfect accuracy of 100\% when predicting between two classes: demented or non-demented. The OASIS dataset, one of the largest datasets, is used for similar 
 experiments with 4-class, 3-class, and 2-class classifications. We achieve very promising performance with accuracies of 99.92\%, 99.90\%, and 99.95\%, respectively.

We also use two ADNI datasets in this study. For the ADNI-1 dataset, we conduct four experiments using our proposed architecture. We create datasets for three different 
brain regions—axial, coronal, and sagittal planes—and one combining all planes. Our architecture performs exceptionally well in all cases. For the axial plane, we achieve
 an accuracy of 99.08\% with an AUC score of 99.80\%. The sagittal plane yields the highest accuracy, with 99.85\% correctly classifying Alzheimer’s progression. For the 
 coronal plane and the mixed-plane dataset, we achieve accuracies of 99.50\% and 99.17\%, respectively. For this dataset, 3-class classification is performed in all 
 experiments.For the ADNI-2 dataset, we conduct two experiments on 5-class and 3-class classifications. For 5-class classification, our proposed method achieves an 
 accuracy of 97.79\%, and for 3-class classification, we achieve 98.60\% accuracy.

When comparing our results with existing studies, we find that most studies using the ADNI and OASIS datasets use significantly smaller sample sizes. In contrast, 
our study achieves superior results and conducts a more extensive and detailed analysis. A comparison of results from existing literature is shown in \autoref{table_summary}, 
and the detailed results of our experiments are presented in \autoref{result}. 

\begin{landscape}
  \centering
  \captionof{table}{Comparison of our Proposed Frameworks with other Recent Works}
  \small
  \setlength\tabcolsep{4pt}
  \begin{longtable}{>{\scriptsize}l l c c c c c c c c} 
  \hline
  \multirow{2}{*}{\textit{ \textbf{Paper}}} & \multirow{2}{*}{\textit{\textbf{Approach}}} & \multirow{2}{*}{\textit{\textbf{Datasets}}} & \multirow{2}{*}{\textit{\textbf{Data Type}}} & \multirow{2}{*}{\textit{\textbf{\makecell[l]{Number of \\Classes}}}} & \multirow{2}{*}{\textit{\textbf{\makecell[l]{Number of\\ Samples}}}} & \multicolumn{4}{c}{\textit{\textbf{Performance}}} \\\cline{7-10}

 & & & & & & \textbf{\textit{Accuracy}} & \textbf{\textit{Precision}}& \textbf{\textit{Recall}}& \textbf{\textit{F1-score}} \\
\hline
  \endfirsthead

\hline

\multirow{2}{*}{\textit{ \textbf{Paper}}} & \multirow{2}{*}{\textit{\textbf{Approach}}} & \multirow{2}{*}{\textit{\textbf{Datasets}}} & \multirow{2}{*}{\textit{\textbf{Data Type}}} & \multirow{2}{*}{\textit{\textbf{\makecell[l]{Number of \\Classes}}}} & \multirow{2}{*}{\textit{\textbf{\makecell[l]{Number of\\ Samples}}}} & \multicolumn{4}{c}{\textit{\textbf{Performance}}} \\ \cline{7-10}
 & & & & & & \textbf{\textit{Accuracy}} & \textbf{\textit{Precision}}& \textbf{\textit{Recall}}& \textbf{\textit{F1-score}} \\
\hline

  \multicolumn{8}{r}{\textit{--continued from previous page}} \\
  \endhead

  \hline
  \multicolumn{8}{r}{\textit{Continued on next page}} \\
  \endfoot

  \endlastfoot
  
\cite{murugan2021demnet} & CNN & Kaggle& MRI& 4 &\makecell{6400 to \\128000 using\\ SMOTE Augment} &95.23\% &96\% &95\% & 95.27\% \\\hline

\multirow{3}{*}{\cite{kaplan2021feed}}& \multirow{3}{*}{\makecell[l]{Local Phase \\Quantization \\ Network}} & \multirow{3}{*}{Kaggle}& \multirow{3}{*}{MRI}&2 & \multirow{3}{*}{-} &99.64\% & 99.72\% &- & 99.64\% \\
& & & &4 & &99.62\% &99.74\% &99.66\% &99.70\% \\ 
&&&&&&&&&\\ \hline

\cite{sharma2022htlml} &\makecell[l]{Transfer Learning + \\Permutation \\ Based Voting} &Kaggle & MRI& 4 & 6126& 91.75\%& -& -& 90.25\%\\\hline

\cite{al2022diagnosis} &AlexNet & Kaggle& MRI& -& -&94.53\% &- &- &94.12\% \\\hline

\cite{ullah2023deep} & CNN& Kaggle& MRI& 4 & \makecell{6400 to \\128000 using\\ SMOTE Augment}& 99.38\%& 99\%& 99\%&99\% \\\hline

\cite{biswas2021enhanced} &CNN  & Kaggle&MRI &2 &4800 & 99.38\%& 99.70\%& 95\%&99.32\% \\\hline

\cite{de2023explainable} &3D CNN & ADNI& \makecell[c]{18F-FDG\\PET}& 2& 2552& 92\%&- & -&- \\\hline

\cite{jahan2023comparison} & \makecell[l]{EfficientNet- \\B7} & ADNI& MRI&5 &10,025 & 96.34\%& 96\%& 96\%&96\% \\\hline

\cite{arafa2024deep} &\makecell[l]{CNN \\VGG16} &Kaggle &MRI & 2&6400 & 97.44\%& 97.46\%& 97.49\%& 97.48\%\\\hline

\cite{noh2023classification} &\makecell[l]{3D-CNN- Phase \\LSTM} &ADNI &fMRI &4 &- &96.43\% &- &- &- \\\hline

\cite{bamber2023medical} & CNN& OASIS& MRI& 4& -& 98\%&- &- & -\\\hline

\multirow{2}{*}{\cite{el2024novel}} & \multirow{2}{*}{CNN}& \multirow{2}{*}{ADNI}& \multirow{2}{*}{MRI}&3 &\multirow{2}{*}{-} &99.43\% &99.43\% &99.43\% &99.43\% \\
& & & & 4& & 99.57\%& 99.57\%&99.57\% & 99.57\%\\\hline

\cite{dhaygude2024knowledge} & 3D-CNN&ADNI &MRI &2 &638 &94.48\% &- &- &- \\\hline

\multirow{2}{*}{\cite{tripathy2024alzheimer}} & \multirow{2}{*}{\makecell[l]{Depthwise \\Separarable CNN }}&\multirow{2}{*}{OASIS} &\multirow{2}{*}{MRI} & \multirow{2}{*}{4}&399 & 96.25\%& 96.71\%& 96.36\%& 96.52\%\\
& & Kaggle& & &6400 &99.75\% &99.63\% &99.77\% & 99.99\%\\
\hline

\multirow{3}{*}{\cite{mahim2024unlocking}} & \multirow{3}{*}{\makecell[l]{Vision \\Transformer + \\ GRU}} & \multirow{2}{*}{Kaggle}&\multirow{3}{*}{MRI} & 4& 6400& 99.53\%& 99.53\%&99.53\% &99.53\% \\
& & & & 2& 6400&99.69\% &99.69\% &99.69\% &99.69\% \\

& & ADNI& & 3&2970 & 99.26\%& 99.26\%&99.26\% &99.26\% \\
\hline

\end{longtable}\label{table_summary}

\end{landscape}

\begin{figure}[H]
  
  \centering
    \graphicspath{./fig/kaggle_conf.png}
  \includegraphics[width=\textwidth, height=0.23\textheight]{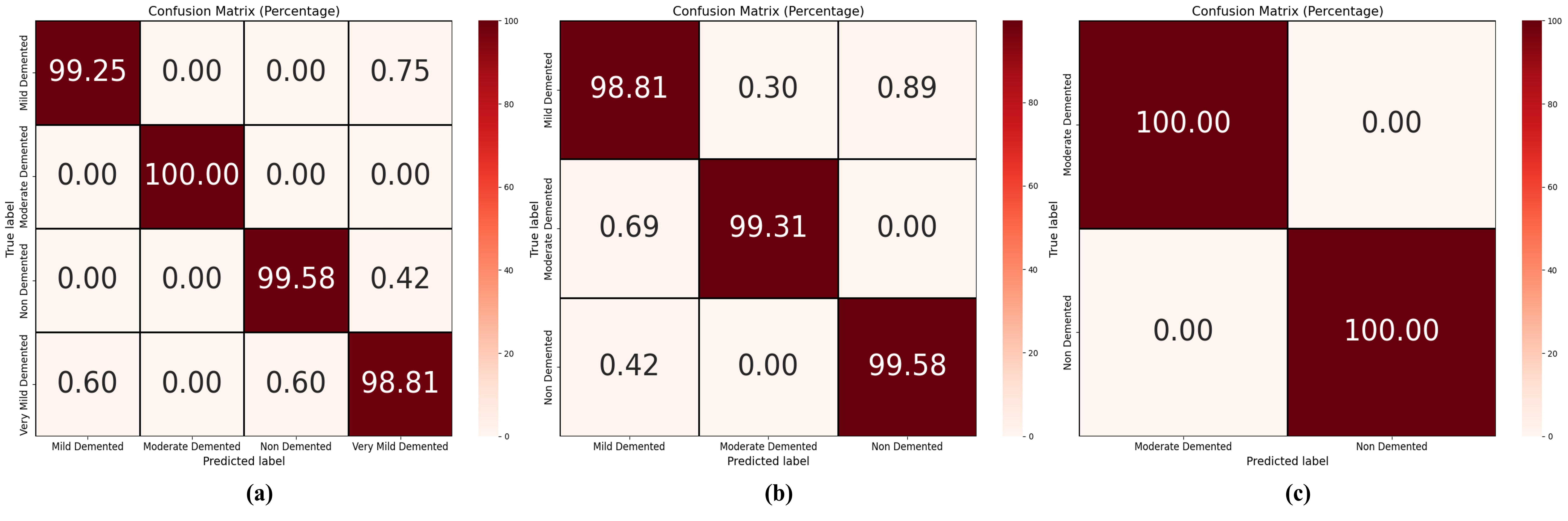}
  
  \caption{Confusion matrix of Kaggle Dataset: (a) 4 Classes (b) 3 classes (c) 2 classes}\label{Kaggle_confusion}
  
\end{figure}

\begin{figure}[H]
  
  \centering
  \graphicspath{./fig/oasis_conf.png}
  \includegraphics[width=\textwidth, height=0.23\textheight]{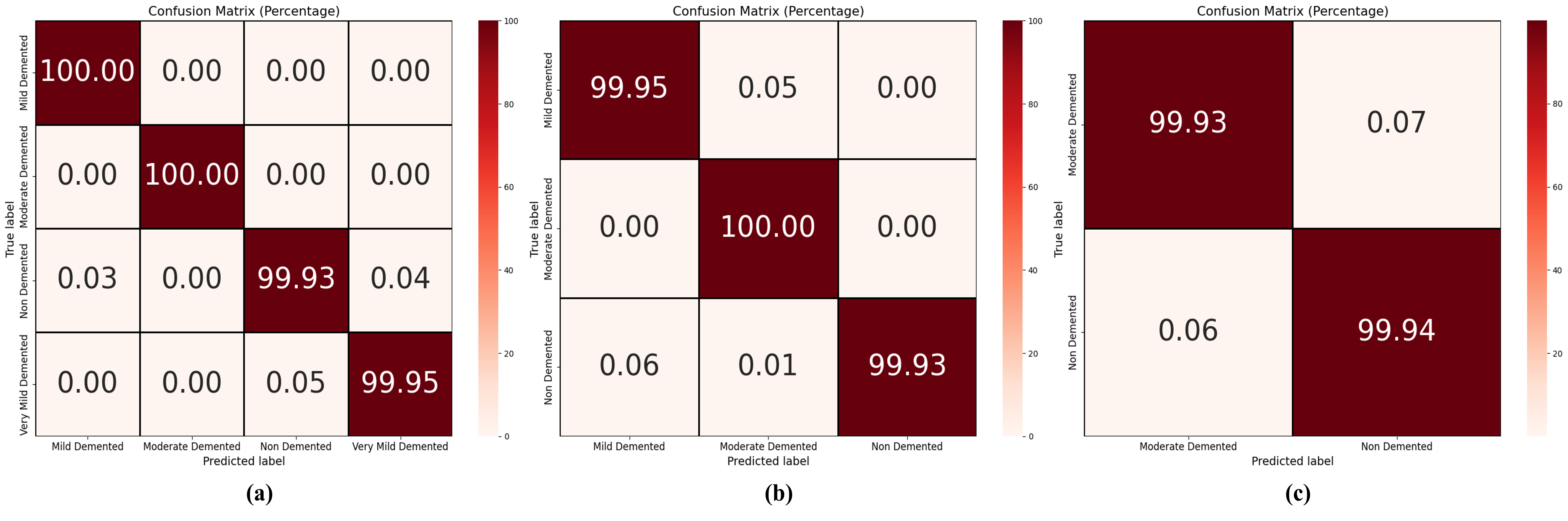}
  
  \caption{Confusion matrix of OASIS Dataset: (a) 4 Classes (b) 3 classes (c) 2 classes}\label{oasis_confusion}
  
\end{figure}

\begin{figure}[H]
  
  \centering
  \graphicspath{./fig/adni_1_confusion.png}
  \includegraphics[width=\textwidth, height=0.18\textheight]{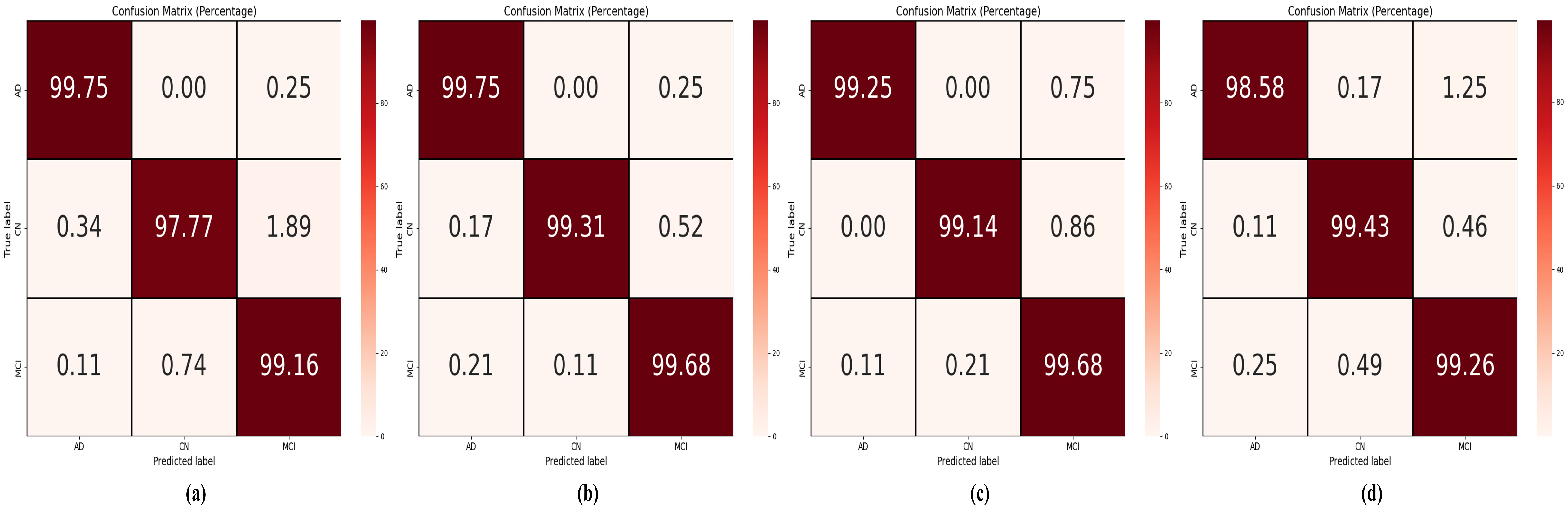}
  
  \caption{Confusion matrix of ADNI-1 Dataset: (a) 3 Classes-axial (b) 3 classes- sagittal (c) 3 classes- coronal (d) 3 classes- all planes}\label{adni_1_confusion}
  
\end{figure}

\begin{figure}[H]
  
  \centering
  \graphicspath{./fig/adni_2_conf.png}
  \includegraphics[width=0.75\textwidth, height=0.23\textheight]{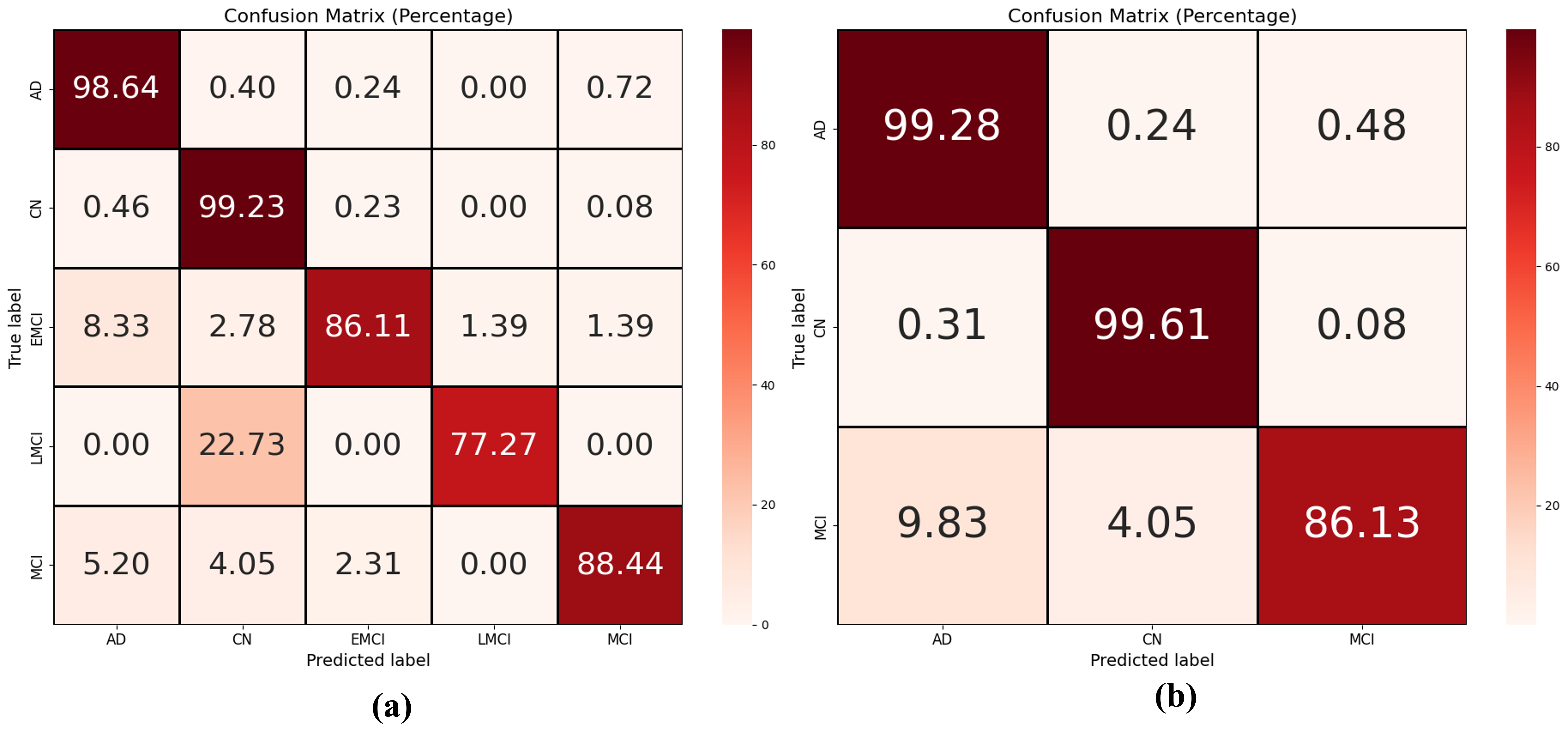}
  
  \caption{Confusion matrix of ADNI-GO Dataset: (a) 5 Classes (b) 3 classes}\label{adni_2_confusion}
  
\end{figure}

\begin{table}[ht]
    \centering
    \small
    \setlength\tabcolsep{2.6pt}
    \caption{Results of our proposed method on four different datasets and with different numbers of classes}
    \begin{tabular}{c c c c c c c c c c}
    \hline
         \multirow{2}{*}{\textit{Dataset}} & \multirow{2}{*}{\textit{Classes}} & \multirow{2}{*}{\textit{Total Samples}} & \multicolumn{6}{c}{\textit{Results}}\\\cline{4-10}
         & & & Accuracy & Precision & Recall & F1-score & RMSE & AUC  & Sensitivity \\
         \hline
         \multirow{3}{*}{\makecell[l]{Kaggle \\Dataset}} & 4 & \multirow{3}{*}{6400} & 99.66\% & 99.66\% & 99.66\% & 99.66\% & 0.0547 & 97.12\% &99.21\%\\ 
         & 3 & & 99.63\% & 99.63\% & 99.63\% & 99.63\%& 0.0157 & 99.81\% &99.24\%\\
         & 2 & & 100\% & 100\% & 100\% & 100\% & 0.000607 & 99.80\% &100\%\\ \hline
         \multirow{3}{*}{\makecell[l]{OASIS}} & 4 & \multirow{3}{*}{80,000} & 99.92\% & 99.92\% & 99.92\% & 99.92\% & 0.0157 &99.14\% & 99.97\%\\
         & 3 & & 99.90\% & 99.90\% & 99.90\% & 99.90\% & 0.0241 & 99.88\% &99.96\% \\
         & 2 & & 99.95\% & 99.95\% & 99.95\% & 99.95\%  & 0.0202 & 99.92\% & 99.9.4\%\\ \hline
         \multirow{2}{*}{\makecell[l]{ADNI-GO}} & 5 & 18,775 & 97.79\% & 97.79\% & 97.79\% & 97.79\% & 0.0927 & 92.71\% & 89.94\% \\
         & 3 & 18,151 & 98.60\% & 98.60\% & 98.60\% & 98.60\% & 0.0967 & 98.23\%& 95\% \\ \hline
         \multirow{4}{*}{\makecell[l]{ADNI-1}} & \makecell[c]{3\\ (Axial Plane)} & \multirow{3}{*}{12,860} & 99.08\% & 99.08\% & 99.08\% & 99.08\% & 0.0634 & 99.80\% &  98.89\%\\
         & \makecell[c]{3\\ (Sagittal Plane)}  & & 99.85\% & 99.85\% & 99.85\% & 99.85\% & 0.0251 &99.91\% & 99.58\% \\
         & \makecell[c]{3\\ (Coronal Plane)} & & 99.5\% & 99.56\% & 99.56\% & 99.56\% & 0.0481 & 99.87\%& 99.35\% \\\cline{2-10}
         & \makecell[c]{3\\ (All Plane)} & 38,580 & 99.17\% & 99.17\% & 99.17\% & 99.17\% & 0.0805 & 99.48\%& 99.08\%\\ \hline

    \end{tabular}
    
    \label{result}
\end{table}
We added visual aids to these numerical data to give a more comprehensive view of our investigations. In the figures below, we present the confusion matrix, 
offering a detailed breakdown of its classification performance. The confusion matrices in Figures \ref{Kaggle_confusion}, \ref{oasis_confusion}, \ref{adni_1_confusion}, and  \ref{adni_2_confusion} 
shows the detailed result overview of all experiments conducted with proposed approach with different classes for datasets: kaggle, oasis, adni-1 and adni-go respectively.
\subsection{Explainability Analysis}
The \autoref{explain} below, highlights areas of significance for decision-making by comparing saliency maps from several models applied to brain MRI scans. 
Concentrated red/yellow regions in the heatmaps show that the suggested model has superior focus, paying closer attention to areas that are diagnostically 
significant. Other models, on the other hand, tend to highlight unimportant regions and show more scattered or chaotic attention. This demonstrates how the 
proposed framework may more effectively capture clinically relevant aspects, improving its interpretability and usefulness in medical applications.
\begin{figure}[H]
  
  \centering
  \graphicspath{./fig/xai.png}
  \includegraphics[width=\textwidth, height=0.8\textheight]{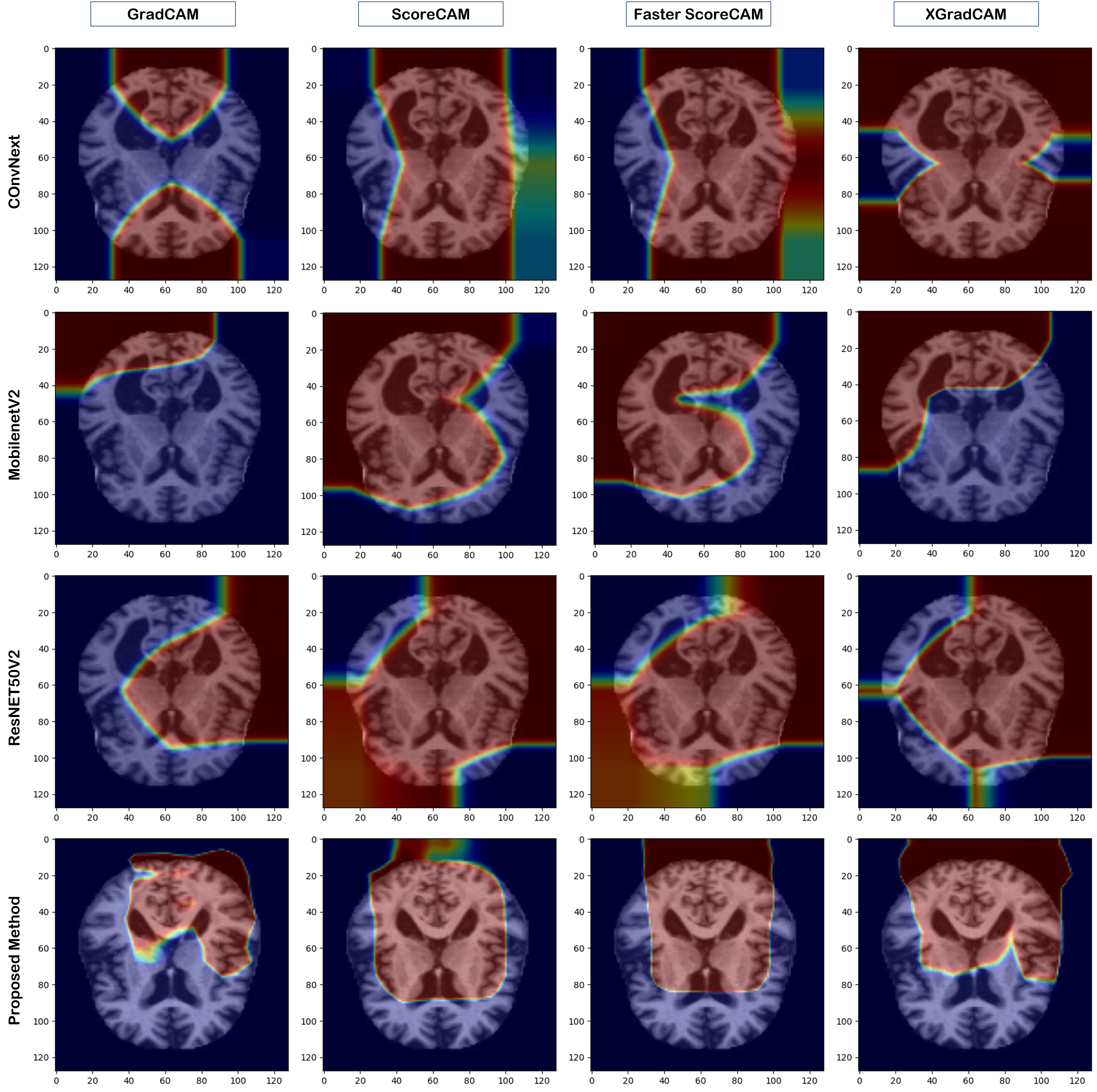}
  
  \caption{Explainability comparison of proposed method with other state-of-the-art methods}\label{explain}
  
\end{figure}

\subsection{Ablation study}
\begin{figure}[H]
  
  \centering
  \graphicspath{./fig/img_preprocessed.png}
  \includegraphics[width=\textwidth, height=0.15\textheight]{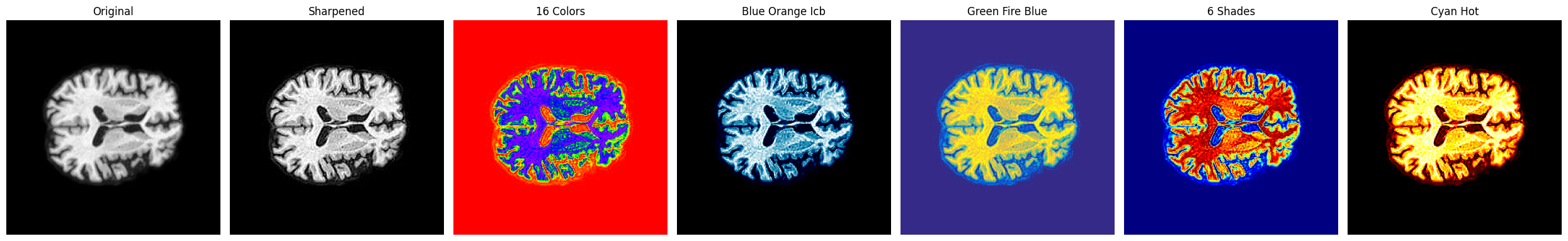}
  
  \caption{Different processed image: original, sharpened, Rainbow,
 Blue Orange Icb, Green Fire Blue, 6 Shades, Cyan Hot }\label{img_proc}
  
\end{figure}

Understanding the contributions of various components to a machine learning model's performance requires comprehensive ablation studies. 
Within the framework of our CNN image categorization model, we conduct a detailed investigation to assess the impact of different hyperparameters and architectural choices. Prior to 
finalizing the hyperparameters, we perform extensive experiments by varying their values. For optimizer selection, we test Adam, SGD, and RMSProp optimizers with 
learning rates of 0.001, 0.005, 0.0001, 0.0005, and 0.00001. For learning rate schedulers, we evaluate ReduceLROnPlateau, Exponential Decay, and Cosine Annealing Decay. 
Based on these experiments, we conclude that the Adam optimizer with an initial learning rate of 0.0001 is the optimal choice. We also explore different image 
preprocessing techniques and data augmentation ranges. For image processing, we experiment with original images, sharpened images, and various colormap applications. 
A sample image of preprocessing techniques is provided in \autoref{img_proc}. Additionally, we test various network architectures by altering the number of layers and 
kernel sizes. To evaluate the model's performance under different conditions, we conduct experiments using both unbalanced datasets and balanced datasets with augmentation. 
These investigations provide insights into the impact of preprocessing, augmentation, and architectural design on model performance.

\section{Conclusion and Future Directions} \label{conclusion}
Effective diagnosis of neurological disorders such as AD and MCI depends on prompt detection and treatments. By presenting a state-of-the-art solution that combines 
the transparency of XAI with the capability of sophisticated deep learning algorithms, the study described here helps to meet this pressing requirement. Our suggested 
architecture is both lightweight and sophisticated. The architecture includes two self-attention layers, the multihead-attention layer and group query attention, 
as well as a complicated spatial attention layer to increase emphasis on spatial and temporal elements. The proposed approach performs better across all four datasets 
and experimentation categories. The model's interpretability is further improved by the use of XAI approaches, which give doctors a better understanding of the 
decision-making process and the factors affecting AD diagnosis. Clinicians may improve treatment strategies and make well-informed judgments by visualizing the important 
features that contribute to the model's predictions. The suggested paradigm has the potential to diagnose AD early and accurately, allowing for swift responses and better 
patient care. Although our findings are encouraging, there is still much work to be done in order to diagnose cognitive diseases as accurately as possible. Subsequent versions 
of our study will explore the possible advantages of using different kernel functions to better combine multimodal data. In order to increase the range of our diagnosis, we are 
also eager to investigate the integration of several imaging modalities. The proposed architecture is highly suited for image-based approaches. For further enhancement, 
incorporating graph theory could offer valuable new insights into AD diagnosis. In conclusion, our research represents a significant step forward in advancing cognitive 
disease detection.

\bibliographystyle{apalike}  

\bibliography{references} 

\begin{thebibliography}{}

\bibitem[AbdulAzeem et~al., 2021]{abdulazeem2021cnn}
AbdulAzeem, Y., Bahgat, W.~M., and Badawy, M. (2021).
\newblock A cnn based framework for classification of alzheimer’s disease.
\newblock {\em Neural Computing and Applications}, 33(16):10415--10428.

\bibitem[Abuhmed et~al., 2021]{abuhmed2021robust}
Abuhmed, T., El-Sappagh, S., and Alonso, J.~M. (2021).
\newblock Robust hybrid deep learning models for alzheimer’s progression
  detection.
\newblock {\em Knowledge-Based Systems}, 213:106688.

\bibitem[Adarsh et~al., 2024]{adarsh2024multimodal}
Adarsh, V., Gangadharan, G., Fiore, U., and Zanetti, P. (2024).
\newblock Multimodal classification of alzheimer's disease and mild cognitive
  impairment using custom mkscddl kernel over cnn with transparent
  decision-making for explainable diagnosis.
\newblock {\em Scientific Reports}, 14(1):1774.

\bibitem[Ahanger et~al., 2024]{ahanger2024alzhinet}
Ahanger, A.~B., Aalam, S.~W., Assad, A., Macha, M.~A., and Bhat, M.~R. (2024).
\newblock Alzhinet: an explainable self-attention based classification model to
  detect alzheimer from 3d volumetric mri data.
\newblock {\em International Journal of System Assurance Engineering and
  Management}, pages 1--10.

\bibitem[Al-Adhaileh, 2022]{al2022diagnosis}
Al-Adhaileh, M.~H. (2022).
\newblock Diagnosis and classification of alzheimer's disease by using a
  convolution neural network algorithm.
\newblock {\em Soft Computing}, 26(16):7751--7762.

\bibitem[Almohimeed et~al., 2023]{almohimeed2023explainable}
Almohimeed, A., Saad, R.~M., Mostafa, S., El-Rashidy, N., Farag, S., Gaballah,
  A., Abd~Elaziz, M., El-Sappagh, S., and Saleh, H. (2023).
\newblock Explainable artificial intelligence of multi-level stacking ensemble
  for detection of alzheimer’s disease based on particle swarm optimization
  and the sub-scores of cognitive biomarkers.
\newblock {\em IEEE Access}.

\bibitem[Ansingkar et~al., 2022]{ansingkar2022efficient}
Ansingkar, N., Patil, R.~B., and Deshmukh, P. (2022).
\newblock An efficient multi class alzheimer detection using hybrid equilibrium
  optimizer with capsule auto encoder.
\newblock {\em Multimedia Tools and Applications}, 81(5):6539--6570.

\bibitem[Arafa et~al., 2024]{arafa2024deep}
Arafa, D.~A., Moustafa, H. E.-D., Ali, H.~A., Ali-Eldin, A.~M., and Saraya,
  S.~F. (2024).
\newblock A deep learning framework for early diagnosis of alzheimer’s
  disease on mri images.
\newblock {\em Multimedia Tools and Applications}, 83(2):3767--3799.

\bibitem[Association, 2019]{alzheimer20192019}
Association, A. (2019).
\newblock 2019 alzheimer's disease facts and figures.
\newblock {\em Alzheimer's \& dementia}, 15(3):321--387.

\bibitem[Bamber and Vishvakarma, 2023]{bamber2023medical}
Bamber, S.~S. and Vishvakarma, T. (2023).
\newblock Medical image classification for alzheimer’s using a deep learning
  approach.
\newblock {\em Journal of Engineering and Applied Science}, 70(1):54.

\bibitem[Bari~Antor et~al., 2021]{bari2021comparative}
Bari~Antor, M., Jamil, A.~S., Mamtaz, M., Monirujjaman~Khan, M., Aljahdali, S.,
  Kaur, M., Singh, P., and Masud, M. (2021).
\newblock A comparative analysis of machine learning algorithms to predict
  alzheimer’s disease.
\newblock {\em Journal of Healthcare Engineering}, 2021(1):9917919.

\bibitem[Beltran et~al., 2020]{beltran2020inexpensive}
Beltran, J.~F., Wahba, B.~M., Hose, N., Shasha, D., Kline, R.~P., and
  Initiative, A. D.~N. (2020).
\newblock Inexpensive, non-invasive biomarkers predict alzheimer transition
  using machine learning analysis of the alzheimer’s disease neuroimaging
  (adni) database.
\newblock {\em PloS one}, 15(7):e0235663.

\bibitem[Biswas et~al., 2021]{biswas2021enhanced}
Biswas, M., Mahbub, M.~K., and Miah, M. A.~M. (2021).
\newblock An enhanced deep convolution neural network model to diagnose
  alzheimer’s disease using brain magnetic resonance imaging.
\newblock In {\em International Conference on Recent Trends in Image Processing
  and Pattern Recognition}, pages 42--52. Springer.

\bibitem[B{\"o}hle et~al., 2019]{bohle2019layer}
B{\"o}hle, M., Eitel, F., Weygandt, M., and Ritter, K. (2019).
\newblock Layer-wise relevance propagation for explaining deep neural network
  decisions in mri-based alzheimer's disease classification.
\newblock {\em Frontiers in aging neuroscience}, 11:456892.

\bibitem[Cole and Franke, 2017]{cole2017predicting}
Cole, J.~H. and Franke, K. (2017).
\newblock Predicting age using neuroimaging: innovative brain ageing
  biomarkers.
\newblock {\em Trends in neurosciences}, 40(12):681--690.

\bibitem[De~Santi et~al., 2023]{de2023explainable}
De~Santi, L.~A., Pasini, E., Santarelli, M.~F., Genovesi, D., and Positano, V.
  (2023).
\newblock An explainable convolutional neural network for the early diagnosis
  of alzheimer’s disease from 18f-fdg pet.
\newblock {\em Journal of Digital Imaging}, 36(1):189--203.

\bibitem[Dhaygude et~al., 2024]{dhaygude2024knowledge}
Dhaygude, A.~D., Ameta, G.~K., Khan, I.~R., Singh, P.~P., Maaliw~III, R.~R.,
  Lakshmaiya, N., Shabaz, M., Khan, M.~A., Hussein, H.~S., and Alshazly, H.
  (2024).
\newblock Knowledge-based deep learning system for classifying alzheimer's
  disease for multi-task learning.
\newblock {\em CAAI Transactions on Intelligence Technology}.

\bibitem[Ding et~al., 2018]{ding2018hybrid}
Ding, X., Bucholc, M., Wang, H., Glass, D.~H., Wang, H., Clarke, D.~H.,
  Bjourson, A.~J., Dowey, L. R.~C., O’Kane, M., Prasad, G., et~al. (2018).
\newblock A hybrid computational approach for efficient alzheimer’s disease
  classification based on heterogeneous data.
\newblock {\em Scientific reports}, 8(1):9774.

\bibitem[Ebrahimi et~al., 2021]{ebrahimi2021convolutional}
Ebrahimi, A., Luo, S., and Disease Neuroimaging~Initiative, f. t.~A. (2021).
\newblock Convolutional neural networks for alzheimer’s disease detection on
  mri images.
\newblock {\em Journal of Medical Imaging}, 8(2):024503--024503.

\bibitem[El-Assy et~al., 2024]{el2024novel}
El-Assy, A., Amer, H.~M., Ibrahim, H., and Mohamed, M. (2024).
\newblock A novel cnn architecture for accurate early detection and
  classification of alzheimer’s disease using mri data.
\newblock {\em Scientific Reports}, 14(1):3463.

\bibitem[El-Sappagh et~al., 2021]{el2021multilayer}
El-Sappagh, S., Alonso, J.~M., Islam, S.~R., Sultan, A.~M., and Kwak, K.~S.
  (2021).
\newblock A multilayer multimodal detection and prediction model based on
  explainable artificial intelligence for alzheimer’s disease.
\newblock {\em Scientific reports}, 11(1):2660.

\bibitem[Ieracitano et~al., 2020]{ieracitano2020novel}
Ieracitano, C., Mammone, N., Hussain, A., and Morabito, F.~C. (2020).
\newblock A novel multi-modal machine learning based approach for automatic
  classification of eeg recordings in dementia.
\newblock {\em Neural Networks}, 123:176--190.

\bibitem[Jahan et~al., 2023a]{jahan2023explainable}
Jahan, S., Abu~Taher, K., Kaiser, M.~S., Mahmud, M., Rahman, M.~S., Hosen,
  A.~S., and Ra, I.-H. (2023a).
\newblock Explainable ai-based alzheimer’s prediction and management using
  multimodal data.
\newblock {\em Plos one}, 18(11):e0294253.

\bibitem[Jahan et~al., 2023b]{jahan2023comparison}
Jahan, S., Saif~Adib, M.~R., Mahmud, M., and Kaiser, M.~S. (2023b).
\newblock Comparison between explainable ai algorithms for alzheimer’s
  disease prediction using efficientnet models.
\newblock In {\em International conference on brain informatics}, pages
  357--368. Springer.

\bibitem[Jain et~al., 2019]{jain2019convolutional}
Jain, R., Jain, N., Aggarwal, A., and Hemanth, D.~J. (2019).
\newblock Convolutional neural network based alzheimer’s disease
  classification from magnetic resonance brain images.
\newblock {\em Cognitive Systems Research}, 57:147--159.

\bibitem[Jin et~al., 2019]{jin2019attention}
Jin, D., Xu, J., Zhao, K., Hu, F., Yang, Z., Liu, B., Jiang, T., and Liu, Y.
  (2019).
\newblock Attention-based 3d convolutional network for alzheimer’s disease
  diagnosis and biomarkers exploration.
\newblock In {\em 2019 IEEE 16Th international symposium on biomedical imaging
  (ISBI 2019)}, pages 1047--1051. IEEE.

\bibitem[Kang et~al., 2023]{kang2023interpretable}
Kang, W., Li, B., Papma, J.~M., Jiskoot, L.~C., Deyn, P. P.~D., Biessels,
  G.~J., Claassen, J.~A., Middelkoop, H.~A., Flier, W. M. v.~d., Ramakers,
  I.~H., et~al. (2023).
\newblock An interpretable machine learning model with deep learning-based
  imaging biomarkers for diagnosis of alzheimer’s disease.
\newblock In {\em International Conference on Medical Image Computing and
  Computer-Assisted Intervention}, pages 69--78. Springer.

\bibitem[Kaplan et~al., 2021]{kaplan2021feed}
Kaplan, E., Dogan, S., Tuncer, T., Baygin, M., and Altunisik, E. (2021).
\newblock Feed-forward lpqnet based automatic alzheimer's disease detection
  model.
\newblock {\em Computers in Biology and Medicine}, 137:104828.

\bibitem[Kavitha et~al., 2022]{kavitha2022early}
Kavitha, C., Mani, V., Srividhya, S., Khalaf, O.~I., and Tavera~Romero, C.~A.
  (2022).
\newblock Early-stage alzheimer's disease prediction using machine learning
  models.
\newblock {\em Frontiers in public health}, 10:853294.

\bibitem[Lahmiri, 2023]{lahmiri2023integrating}
Lahmiri, S. (2023).
\newblock Integrating convolutional neural networks, knn, and bayesian
  optimization for efficient diagnosis of alzheimer's disease in magnetic
  resonance images.
\newblock {\em Biomedical Signal Processing and Control}, 80:104375.

\bibitem[Lee et~al., 2019a]{lee2019using}
Lee, B., Ellahi, W., and Choi, J.~Y. (2019a).
\newblock Using deep cnn with data permutation scheme for classification of
  alzheimer's disease in structural magnetic resonance imaging (smri).
\newblock {\em IEICE TRANSACTIONS on Information and Systems},
  102(7):1384--1395.

\bibitem[Lee et~al., 2019b]{lee2019predicting}
Lee, G., Nho, K., Kang, B., Sohn, K.-A., and Kim, D. (2019b).
\newblock Predicting alzheimer’s disease progression using multi-modal deep
  learning approach.
\newblock {\em Scientific reports}, 9(1):1952.

\bibitem[Liu et~al., 2022]{liu2022diagnosis}
Liu, Z., Lu, H., Pan, X., Xu, M., Lan, R., and Luo, X. (2022).
\newblock Diagnosis of alzheimer’s disease via an attention-based multi-scale
  convolutional neural network.
\newblock {\em Knowledge-Based Systems}, 238:107942.

\bibitem[Mahim et~al., 2024]{mahim2024unlocking}
Mahim, S., Ali, M.~S., Hasan, M.~O., Nafi, A. A.~N., Sadat, A., Al~Hasan, S.,
  Shareef, B., Ahsan, M.~M., Islam, M.~K., Miah, M.~S., et~al. (2024).
\newblock Unlocking the potential of xai for improved alzheimer’s disease
  detection and classification using a vit-gru model.
\newblock {\em IEEE Access}.

\bibitem[Moser et~al., 2009]{moser2009magnetic}
Moser, E., Stadlbauer, A., Windischberger, C., Quick, H.~H., and Ladd, M.~E.
  (2009).
\newblock Magnetic resonance imaging methodology.
\newblock {\em European journal of nuclear medicine and molecular imaging},
  36:30--41.

\bibitem[Murugan et~al., 2021]{murugan2021demnet}
Murugan, S., Venkatesan, C., Sumithra, M., Gao, X.-Z., Elakkiya, B., Akila, M.,
  and Manoharan, S. (2021).
\newblock Demnet: A deep learning model for early diagnosis of alzheimer
  diseases and dementia from mr images.
\newblock {\em Ieee Access}, 9:90319--90329.

\bibitem[Neugroschl and Wang, 2011]{neugroschl2011alzheimer}
Neugroschl, J. and Wang, S. (2011).
\newblock Alzheimer's disease: diagnosis and treatment across the spectrum of
  disease severity.
\newblock {\em Mount Sinai Journal of Medicine: A Journal of Translational and
  Personalized Medicine}, 78(4):596--612.

\bibitem[Noh et~al., 2023]{noh2023classification}
Noh, J.-H., Kim, J.-H., and Yang, H.-D. (2023).
\newblock Classification of alzheimer’s progression using fmri data.
\newblock {\em Sensors}, 23(14):6330.

\bibitem[Park et~al., 2023]{park2023development}
Park, H.~Y., Shim, W.~H., Suh, C.~H., Heo, H., Oh, H.~W., Kim, J., Sung, J.,
  Lim, J.-S., Lee, J.-H., Kim, H.~S., et~al. (2023).
\newblock Development and validation of an automatic classification algorithm
  for the diagnosis of alzheimer’s disease using a high-performance
  interpretable deep learning network.
\newblock {\em European Radiology}, 33(11):7992--8001.

\bibitem[Petersen et~al., 2010]{petersen2010alzheimer}
Petersen, R.~C., Aisen, P.~S., Beckett, L.~A., Donohue, M.~C., Gamst, A.~C.,
  Harvey, D.~J., Jack~Jr, C., Jagust, W.~J., Shaw, L.~M., Toga, A.~W., et~al.
  (2010).
\newblock Alzheimer's disease neuroimaging initiative (adni) clinical
  characterization.
\newblock {\em Neurology}, 74(3):201--209.

\bibitem[Pinamonti, 2021]{kaggle_data}
Pinamonti, M. (2021).
\newblock {Alzheimer MRI 4 classes dataset}.
\newblock
  \url{https://www.kaggle.com/datasets/marcopinamonti/alzheimer-mri-4-classes-dataset/data}.
\newblock [Online; accessed 20-July-2024].

\bibitem[Rallabandi et~al., 2020]{rallabandi2020automatic}
Rallabandi, V.~S., Tulpule, K., Gattu, M., Initiative, A. D.~N., et~al. (2020).
\newblock Automatic classification of cognitively normal, mild cognitive
  impairment and alzheimer's disease using structural mri analysis.
\newblock {\em Informatics in Medicine Unlocked}, 18:100305.

\bibitem[Salehi et~al., 2020]{salehi2020cnn}
Salehi, A.~W., Baglat, P., Sharma, B.~B., Gupta, G., and Upadhya, A. (2020).
\newblock A cnn model: earlier diagnosis and classification of alzheimer
  disease using mri.
\newblock In {\em 2020 International Conference on Smart Electronics and
  Communication (ICOSEC)}, pages 156--161. IEEE.

\bibitem[Scheltens et~al., 2021]{scheltens2021alzheimer}
Scheltens, P., De~Strooper, B., Kivipelto, M., Holstege, H., Ch{\'e}telat, G.,
  Teunissen, C.~E., Cummings, J., and van~der Flier, W.~M. (2021).
\newblock Alzheimer's disease.
\newblock {\em The Lancet}, 397(10284):1577--1590.

\bibitem[Shahbaz et~al., 2019]{shahbaz2019classification}
Shahbaz, M., Ali, S., Guergachi, A., Niazi, A., and Umer, A. (2019).
\newblock Classification of alzheimer's disease using machine learning
  techniques.
\newblock In {\em International Conference on Data Technologies and
  Applications}.

\bibitem[Shamrat et~al., 2023]{shamrat2023alzheimernet}
Shamrat, F. J.~M., Akter, S., Azam, S., Karim, A., Ghosh, P., Tasnim, Z.,
  Hasib, K.~M., De~Boer, F., and Ahmed, K. (2023).
\newblock Alzheimernet: An effective deep learning based proposition for
  alzheimer’s disease stages classification from functional brain changes in
  magnetic resonance images.
\newblock {\em IEEE Access}, 11:16376--16395.

\bibitem[Sharma et~al., 2022]{sharma2022htlml}
Sharma, S., Gupta, S., Gupta, D., Altameem, A., Saudagar, A. K.~J., Poonia,
  R.~C., and Nayak, S.~R. (2022).
\newblock Htlml: Hybrid ai based model for detection of alzheimer’s disease.
\newblock {\em Diagnostics}, 12(8):1833.

\bibitem[Tripathy et~al., 2024]{tripathy2024alzheimer}
Tripathy, S.~K., Nayak, R.~K., Gadupa, K.~S., Mishra, R.~D., Patel, A.~K.,
  Satapathy, S.~K., Bhoi, A.~K., and Barsocchi, P. (2024).
\newblock Alzheimer’s disease detection via multiscale feature modelling
  using improved spatial attention guided depth separable cnn.
\newblock {\em International Journal of Computational Intelligence Systems},
  17(1):113.

\bibitem[Ullah and Jamjoom, 2023]{ullah2023deep}
Ullah, Z. and Jamjoom, M. (2023).
\newblock A deep learning for alzheimer’s stages detection using brain
  images.
\newblock {\em Computers, Materials \& Continua}, 74(1).

\bibitem[Wang et~al., 2024]{wang2024multimodal}
Wang, C., Tachimori, H., Yamaguchi, H., Sekiguchi, A., Li, Y., Yamashita, Y.,
  and Initiative, A. D.~N. (2024).
\newblock A multimodal deep learning approach for the prediction of cognitive
  decline and its effectiveness in clinical trials for alzheimer’s disease.
\newblock {\em Translational psychiatry}, 14(1):105.

\bibitem[Wu et~al., 2022]{wu2022attention}
Wu, Y., Zhou, Y., Zeng, W., Qian, Q., and Song, M. (2022).
\newblock An attention-based 3d cnn with multi-scale integration block for
  alzheimer's disease classification.
\newblock {\em IEEE Journal of Biomedical and Health Informatics},
  26(11):5665--5673.

\bibitem[Zhang et~al., 2012]{zhang2012multi}
Zhang, D., Shen, D., Initiative, A. D.~N., et~al. (2012).
\newblock Multi-modal multi-task learning for joint prediction of multiple
  regression and classification variables in alzheimer's disease.
\newblock {\em NeuroImage}, 59(2):895--907.

\bibitem[Zhang et~al., 2021a]{zhang20213d}
Zhang, J., Zheng, B., Gao, A., Feng, X., Liang, D., and Long, X. (2021a).
\newblock A 3d densely connected convolution neural network with
  connection-wise attention mechanism for alzheimer's disease classification.
\newblock {\em Magnetic Resonance Imaging}, 78:119--126.

\bibitem[Zhang et~al., 2021b]{zhang2021explainable}
Zhang, X., Han, L., Zhu, W., Sun, L., and Zhang, D. (2021b).
\newblock An explainable 3d residual self-attention deep neural network for
  joint atrophy localization and alzheimer’s disease diagnosis using
  structural mri.
\newblock {\em IEEE journal of biomedical and health informatics},
  26(11):5289--5297.

\bibitem[Zhang and Yang, 2021]{zhang2021survey}
Zhang, Y. and Yang, Q. (2021).
\newblock A survey on multi-task learning.
\newblock {\em IEEE transactions on knowledge and data engineering},
  34(12):5586--5609.

\end{thebibliography}


\end{document}